\documentclass[%
 aip,
 amsmath,amssymb,
 reprint,%
]{revtex4-1}

\usepackage{graphicx}
\usepackage{dcolumn}
\usepackage{bm}

\usepackage[utf8]{inputenc}
\usepackage[T1]{fontenc}
\usepackage{mathptmx}

\usepackage{xcolor}

\graphicspath{{figures/}}

\makeatother
\begin{document}

\preprint{AIP/123-QED}

\title[Effect of collisions on non-adiabatic electron dynamics in ITG-driven microturbulence]{Effect of collisions on non-adiabatic electron dynamics in ITG-driven microturbulence}

\author{Ajay C. J.}%
\email{trax.42@hotmail.com}
\author{Stephan Brunner}
\author{Justin Ball}
\affiliation{ 
Ecole Polytechnique F\'ed\'erale de Lausanne (EPFL), Swiss Plasma Center, CH-1015 Lausanne, Switzerland
}%

\makeatletter
\def\@email#1#2{%
 \endgroup
 \patchcmd{\titleblock@produce}
  {\frontmatter@RRAPformat}
  {\frontmatter@RRAPformat{\produce@RRAP{*#1\href{mailto:#2}{#2}}}\frontmatter@RRAPformat}
  {}{}
}%

\date{\today}

\begin{abstract}
Non-adiabatic electron response leads to significant changes in Ion Temperature Gradient (ITG) eigenmodes, leading in particular to fine-structures that are significantly extended along the magnetic field lines at corresponding Mode Rational Surfaces (MRSs). These eigenmodes can nonlinearly interact with themselves to drive zonal flows via the so-called self-interaction mechanism. In this paper, the effect of collisions on these processes are studied. In presence of non-adiabatic electrons, the linear growth rate of ITG eigenmodes decreases with increasing collisionality. Detailed velocity space analysis of the distribution function shows that this results from collisions leading to a more adiabatic-like response of electrons away from MRSs. In linear simulations, collisions are furthermore found to broaden the radial width of the fine-structures, which translates to narrower tails of the eigenmode in extended ballooning space. The characteristic parallel scale length associated to these tails is shown to scale with the mean free path of electrons. In nonlinear turbulence simulations accounting for physically relevant values of collisionality, the fine structures located at MRSs, together with the associated drive of zonal flows via self-interaction, are shown to persist and play a significant role.
\end{abstract}

\maketitle

\section{Introduction}\label{sec_collisions}
The temperatures and densities typical of tokamak cores lead to plasmas with very low collisionalities, with collision frequencies lower than that of typical frequencies of microinstabilities. Hence these plasmas are often approximated as collisionless and modelled without collisions in gyrokinetic simulations. However, it is important to model collisions properly in the core for various reasons. For instance, collisions are necessary to smooth the small-scale structures in velocity space, which have been observed in gyrokinetic simulations \cite{Watanabe2004,Tatsuno2009}, and play an important role in the energy transfer mechanism \cite{Schekochihin2008}. In fact collisions provide the physical link between macroscopic plasma heating and microturbulence, through dissipation of small-scale structures in both position and velocity space, thereby enabling the system to reach the correct statistical steady state \cite{Krommes2000_2,Abel2008}. 

Collisions can affect the steady-state turbulence levels either by affecting the linear microinstability drive of turbulence or by affecting the saturation mechanism. An example of the former case is the collisional stabilisation of the Trapped Electron Mode (TEM) microinstability, observed even in experiments \cite{Camenen2007}, and sometimes leading to the transition of TEM dominant turbulence to Ion Temperature Gradient (ITG) dominant one with increasing collisionality \cite{Ryter2005}. And an example of collisions affecting turbulence saturation is the collisional damping of zonal flows which in turn can lead to larger heat and particle steady-state flux levels \cite{Lin1998,Hinton1999}.

Earlier studies \cite{Kauffmann2010,VernayPhD} where electrons were modelled to respond adiabatically have reported that collisions do not significantly alter ITG eigenmode growth rates. Note that the adiabatic electron response is valid in the limit of the (parallel) phase velocity [=$\omega/k_{\parallel}$, with $\omega$ being the real frequency of the eigenmode and $k_\parallel$ being the parallel wavenumber] of the wave being much less than the thermal velocity of electrons. However at radial locations corresponding to the Mode Rational Surfaces (MRSs) of an eigenmode, where $k_\parallel\rightarrow 0$, non-adiabatic electron response becomes important. In particular, non-adiabatic passing electron response has been shown to lead to 'fine-structures' in the electrostatic potential, temperature and density perturbations at the MRSs of the respective eigenmode~\cite{Chowdhury2008,Waltz2006,Dominski2015}. Furthermore, the recently studied zonal flow driving mechanism referred to as the self-interaction mechanism~\cite{Weikl2018,AjayCJ2020} is particularly dominant in presence of non-adiabatic electrons and has been shown to possibly play a significant role in determining the steady state flux levels in nonlinear gyrokinetic simulations~\cite{Justin2020,AjayCJ2020}.

In this paper, we focus on the effect of collisions on ITG turbulence in presence of non-adiabatic electron response. The previous work by Mikkelsen \cite{Mikkelsen2008} has already reported that the ITG eigenmodes  are in fact significantly affected by collisions when electrons are treated kinetically. This study is followed up here by exploring the effects of collisions, in particular on the fine-structures associated to non-adiabatic passing electrons and the self-interaction mechanism. 

Using \emph{linear} simulations, through a scan in collisionality, two preliminary results are obtained: 1) growth rate of ITG eigenmode can indeed decrease significantly with increasing collisionality in collisionality regimes typical of the core, and 2), the radial width of fine-structure associated to non-adiabatic passing electron response broadens with increasing collisionality. While the first result has already been reported in Ref.~\onlinecite{Mikkelsen2008}, its fundamental reason has not been illustrated in detail. In this paper, through a detailed velocity space analysis of the distribution function, it is shown that collisions lead to a more adiabatic-like response of electrons away from MRSs, which in turn explains the decrease in growth rates with increasing collisionality. Furthermore it is shown that collisionality sets the characteristic parallel length scale associated to the ballooning envelope tail of eigenmodes, which in turn explains the radial broadening of the fine-structures.

In \emph{nonlinear} turbulence simulations, the eigenmodes can get deformed by the various nonlinear mechanisms \cite{AjayCJ2020}. Furthermore, as already mentioned, collisions can affect the nonlinear turbulence saturation mechanism through damping of zonal flows. Of particular interest to this work is investigating how collisions affect the nonlinear drive of zonal flows via the self-interaction mechanism. It is found that the steady-state heat flux decreases with increasing collisionality which is then illustrated to be the consequence of the corresponding decrease in the growth rate of linear eigenmodes. The width of the fine-structures in nonlinear simulations is however not found to show an increase with increasing collisionality, in the contrary. This is a consequence of the dominant nonlinear broadening effect of these fine structures, which in fact decreases due to the reduced instability drive with increasing collisionality. Finally the effect of collisions on the self-interaction mechanism is studied using the diagnostic methods developed earlier in Ref.~\onlinecite{AjayCJ2020}, more specifically, the time-averaged ballooning structure, time evolution of linear phase difference along the ballooning structure, the normalised self-interaction contribution to Reynolds stress, the bicoherence estimate and the correlation between the various toroidal mode contributions to Reynolds stress. 

The rest of the paper is organised as follows. First, the simulation setup is described in section~\ref{CollSimsetup}. In section~\ref{SecCollLin}, the linear simulation study is presented, in two parts. The analysis on the dependence of growth rate on collisionality and the analysis on the increase in the radial width of fine-structures with collisionality is discussed in sections~\ref{SecCollLinGrowthRate} and \ref{SecCollLinRadialwidth} respectively. The results on the effect of collisions in nonlinear simulations is presented in three subsections which are as follows: The effect of collisions on the heat flux and the shearing rate associated with zonal flows is discussed in section~\ref{SecCollNLflux}, followed by its effect on the radial width of non-linear fine-structures in section~\ref{SecCollNonlinFSWidth}. In section~\ref{SecCollNLSI}, the effect of collisions on the self-interaction mechanism is studied using the diagnostic methods of Ref.~\onlinecite{AjayCJ2020}, also briefly mentioned in the previous paragraph. Finally, the conclusions are presented in section~\ref{SecCollConclusions}.

\section{Simulation setup}\label{CollSimsetup} 

\subsubsection*{Flux-tube gyrokinetic  model and the coordinate system}

The flux-tube version~\cite{Beer1995} of the Eulerian gyrokinetic code GENE~\cite{GENE1,GENE2,GENE3} is used in this study. It considers a field-aligned coordinate system with $x\in[-L_x/2,L_x/2[$ being the radial coordinate, $y\in[0,L_y[$ being the binormal coordinate and $z\in[-\pi,\pi[$ being the parallel coordinate. In the following, a brief description of the flux-tube model and the coordinate system are given. For a more detailed description, including that of the boundary conditions, refer Ref.~\onlinecite{AjayCJ2020}. 

In the flux-tube model, the background density and temperature profiles and their gradients, as well as the magnetic equilibrium quantities, are considered constant across the radial extension $L_x$ of the simulation box, and are evaluated at a radial position denoted by $r_0$. An exception is the safety factor $q_s\simeq q_0(1 + \hat{s}x/r_0)$ which is assumed to have a linear variation across the flux-tube, with $q_0=q_s\rvert_{r=r_0}$, a constant magnetic shear $\hat{s}=(r/q_s)(dq_s/dr)\rvert_{r=r_0}$ and $x=r-r_0$. $r$ has units of length and labels magnetic surfaces, providing an estimate of the (average) minor radius. The background density and temperature of a species $j$ are, respectively, $n_{j,0}=n_{j,0}(r_0)$ and $T_{j,0}=T_{j,0}(r_0)$ and their inverse radial gradient lengths are $1/L_{Nj}=-d~\text{log}~n_{j,0}/dr|_{r=r_0}$ and $1/L_{Tj}=-d~\text{log}~T_{j,0}/dr|_{r=r_0}$.

In the GENE flux-tube model, both the radial and binormal coordinates are treated in Fourier space with the corresponding wavenumbers being $k_x$ and $k_y$ respectively. Given the axisymmetric toroidal geometry of the unperturbed system and the associated parallel boundary condition, a linear eigenmode which has a fixed $k_y$ wavenumber along $y$ involves a set of linearly coupled $k_x$ Fourier modes with $k_x = k_{x0} + p\,2\pi k_y\hat{s}$, $p\in\mathbb{Z}$. An eigenmode is thus of the form:
\begin{equation}
  \label{linear eigenmode, Fourier rep.}
  \mathcal{A}(x, y, z) 
  = 
  e^{ik_yy}
  \sum_{p=-\infty}^{+\infty}
  \hat{\mathcal{A}}_{k_{x0} + p2\pi k_y\hat{s},\,k_y}(z)
  \exp[i(k_{x0} + p\,2\pi k_y\hat{s})x].
\end{equation}
The ballooning representation of the eigenmode is defined as follows ~\cite{Merlo2016}:
\begin{flalign}
  \label{ballooning defs.}
  \hat{\mathcal{A}}_b(z+p\,2\pi) 
  & = 
  \hat{\mathcal{A}}_{k_{x0} + p2\pi k_y\hat{s},\,k_y}(z),
  \hspace{1cm} z\in[-\pi,\pi[
  \end{flalign}
\begin{flalign}
\label{kx0deff}
  \chi_0 
  & = 
  -k_{x0}/(k_y\hat{s}),
\end{flalign}
where the ballooning envelope $\hat{\mathcal{A}}_b(\chi)$ is defined over the extended ballooning space $\chi\in]-\infty, +\infty[$ and $\chi_0$ is the so-called ballooning angle.
.

In a flux-tube of radial extension $L_x$, all coupled Fourier modes $k_x +
p\,2\pi k_y \hat{s}$ relative to this direction must be harmonics of $k_{x,
  \min}=2\pi/L_x$. This must hold for all $k_y$ and in particular for the lowest
harmonic $k_{y, min}$, which implies:
\[
2\pi k_{y, \min} \hat{s} 
= 
M\,k_{x, min} 
= M\,\frac{2\pi}{L_x},
\]
with $M \in \mathbb{N}^\star$ a strictly positive integer. This relation can
be rewritten:
\begin{equation}
  \label{relation between Lx and Ly}
  L_x 
  = 
  \frac{M}{k_{y, \min}\hat{s}}
  = 
  M\,\Delta x_{\rm LMRS} 
  = 
  \frac{M}{2\pi\hat{s}}\,L_y,
\end{equation}
thus imposing a constraint between the extensions $L_x$ and $L_y$ of the flux
tube along the directions $x$ and $y$ respectively. In practice, the integer
$M$ must be chosen such that $L_x$ is larger then the radial correlation
length of turbulent eddies.
  
Relation (\ref{relation between Lx and Ly}) also implies that $L_x$ must be an
integer multiple of $\Delta x_{\rm LMRS} = 1/(k_{y, \min}\hat{s})$, identified
as the distance between lowest order MRSs. Indeed, considering the linearised
safety factor profile, the distance $\Delta
x_{\rm MRS}$ between MRSs corresponding to a given $k_y\neq 0$ mode is
constant and given by
\[
\Delta x_{\rm MRS}(k_y) = 1/(k_y\hat{s}).
\] 
One thus in particular has
$\Delta x_{\rm LMRS} = \Delta x_{\rm MRS}(k_{y,\min}) = 1/(k_{y,
  \min}\hat{s})$.  For a given $k_y\neq 0$ mode, the radial positions of
corresponding MRSs are thus
\[
x_{\rm MRS} 
= 
m\,\Delta x_{\rm MRS} 
= 
m\,\frac{k_{y, \min}}{k_y}\,\Delta x_{\rm LMRS}, 
\quad m\in\mathbb{Z}.
\]

The discretisation of the phase space coordinates is as follows. In real space, the simulation volume
$L_x\times L_y\times L_z$ is discretised by $N_x, N_y$ and $N_z$
equidistant grid points along the directions $x, y$ and $z$ respectively. For the parallel velocity coordinate, one considers $v_\parallel \in [-v_{\parallel, \max}, +v_{\parallel,
\max}]$ with a discretisation involving $N_{v_\parallel}$ equidistant grid
points, while for the magnetic moment coordinate one considers $\mu \in [0,
\mu_{\max}]$ with a discretisation involving $N_\mu$ grid points following the
Gauss-Laguerre integration scheme.

\subsubsection*{Collision frequency}

A brief description on implementation of collisions in GENE is given here. For a more detailed description, see the Refs.~\onlinecite{GENE3,DoerkPhD,CrandallPhD}. 

In GENE, the collision frequency is set via the normalised quantity $\nu_c$ defined as 
\begin{align}
\label{nucdef}
\nu_c=\frac{\pi {\rm ln}\Lambda e^4 n_{\rm ref}L_{\rm ref}}{2^{3/2}T_{\rm ref}^2},
\end{align}
where $n_{\rm ref}$, $L_{\rm ref}$ and $T_{\rm ref}$ are the reference density, equilibrium length and temperature respectively. $e$ is the elementary electron charge and ${\rm ln}\Lambda$ is the Coulomb logarithm. For the simulations considered in this paper, one has $n_{\rm ref}=n_{0,i}=n_{0,e}$, $L_{\rm ref}=R$ the major radius on axis of the tokamak and $T_{\rm ref}=T_{0,i}=T_{0,e}$.  $\nu_c$ is related to the electron-ion collision rate $\nu_{ei}$ \cite{Hinton1976} as
\begin{align}
\label{nueidef}
\nu_{ei}(v)
=\frac{\sqrt{2}\pi Z^2e^4n_{0,i}{\rm ln}\Lambda}{\sqrt{m_e}T_{0,e}^{3/2}}\frac{v_{T,e}^3}{v^3}
=4Z^2\frac{n_{0,i}}{n_{\rm ref}}\frac{T_{\rm ref}^2}{T_{0,e}^2}\frac{v_{T,e}^3}{v^3}\frac{v_{th,e}}{L_{\rm ref}}\nu_c,
\end{align}
where $v_{T,e}=\sqrt{2T_{0,e}/m_e}$. $m_{e/i}$ denotes the mass of electrons/ions respectively. Note that the thermal collision rate is obtained for $v=v_{T,e}$.

A physically more illustrative measure of collision frequency is the normalised collisionality $\nu_e^*$, which estimates the average number of times a trapped electron is scattered to become passing before completing a banana orbit. Naturally, the banana (so-called collisionless) regime for electrons is therefore characterised by $\nu_e^*< 1$ \cite{Hinton1976}. For a general geometry, $\nu_e^*$ is defined as  
\begin{align}
\label{nustardef}
\nu_e^*=\frac{\sqrt{2}a B_0}{B_{p0}v_{T,e}\epsilon^{3/2}\tau_e},
\end{align}
where $a$ is the minor radius of the tokamak, $B_{p0}$ is the poloidal component of the background magnetic field strength, $\epsilon=r/R_0$ is the inverse aspect ratio and $\tau_e=3m_e^2v_{T,e}^3/16\sqrt{\pi}Z^2e^4n_{0,i}{\rm ln}\Lambda$ is the electron-ion momentum exchange time, also called the electron collision time. For the circular ad-hoc geometry considered in this paper, one has $B_{p0}=B_0r_0/(Rq_0)$ and
\begin{align}
\nu_e^*=\frac{16}{3\sqrt{\pi}}\frac{q_0Z^2}{\epsilon^{3/2}}\frac{R}{L_{\rm ref}}\frac{n_i}{n_{\rm ref}}\frac{T_{\rm ref}^2}{T_{0,e}^2}\nu_c.
\end{align}
The corresponding ion collisionality $\nu_i^*$ is given by 
\begin{align}
\nu_i^*=\frac{8}{3\sqrt{\pi}}\frac{q_0Z^2}{\epsilon^{3/2}}\frac{R}{L_{\rm ref}}\frac{n_i}{n_{\rm ref}}\frac{T_{\rm ref}^2}{T_{0,i}^2}\nu_c.
\end{align}

In this study, collisions are modelled with the linearized Landau collision operator, and a scan in collisionality is carried out over the range $0\leq\nu_e^*\leq 2.758$. Note that the background densities $n_0$ and temperatures $T_0$ typical of tokamak cores lead to collisionalities ($\nu^*\propto n_0/T_0^2$) that fall within the banana regime characterised by $\nu^*<1$, while those at the plasma edge can extend towards the plateau regime characterised by $1<\nu^*<\epsilon^{-3/2}$ \cite{Hinton1976}. The Finite Larmor Radius (FLR) corrections to the collision operator leading in particular to spatial diffusion terms in gyrocenter coordinates \cite{Abel2008} are absent in the collision operator which has been used in the simulations shown in this paper. However, by explicitly turning on the FLR contributions, it has been verified that it leads to an insignificant change in the linear simulations results, for the considered range of collisionality.
  
The other parameters used in this study are close to the Cyclone base case~\cite{Dimits2000}, and are given in table~\ref{CollParameterSet}. $\beta$ denotes the ratio of the magnetic pressure to plasma pressure.

\begin{table}
\caption{Parameter set for linear simulations. In parenthesis, parameters for non-linear simulations are given if they differ from those in the linear case.}
\begin{ruledtabular}
\hspace{-0.3cm}
\begin{footnotesize}
\begin{tabular}{ p{2.95cm} p{3.05cm} p{2.5cm}}
  \multicolumn{3}{l}{Geometry: Ad-hoc concentric circular
    geometry \cite{Lapillonne2009}} \vspace{0.1cm}
  \\ 
  $\epsilon=0.18$, & $q_0=1.4$, & $\hat{s}=0.8$, \vspace{0.2cm}
  \\ 
  $m_i/m_e=1836$, & $T_{e}/T_{i}=1.0$, & $\beta=0.001$, \vspace{0.2cm}
  \\
  $R/L_{N}=2.0$, & $R/L_{T_{i}}=8.0$, & $R/L_{T_{e}}=2.0$, \vspace{0.15cm}
  \\ 
  $L_x=3.6\,\rho_i\ (142.9\,\rho_i$), & $L_y=18.0\,\rho_i\ (179.5\,\rho_i)$, & $L_z = 2\pi$, \vspace{0.2cm} 
  \\
  $k_{y,{\rm min}}\rho_i=0.35\hspace{0.05cm}(0.035)$, & $v_{\parallel, \max} = 3\,v_{{T}, i}$, & $\mu_{\max}\hspace{-0.02cm}=\hspace{-0.02cm}9\,T_{0i}/B_{0, {\rm axis}}$, \vspace{0.2cm}
  \\ 
  \multicolumn{2}{l}{$N_{k_x}\hspace{-0.1cm} \times\hspace{-0.04cm} N_{k_y}^* \hspace{-0.1cm}\times\hspace{-0.04cm} N_z \hspace{-0.1cm}\times\hspace{-0.04cm}
    N_{v_{\parallel}} \hspace{-0.1cm}\times\hspace{-0.04cm} N_\mu = 96\hspace{0.02cm}(256) \hspace{-0.05cm}\times\hspace{-0.05cm} 1\hspace{0.02cm}(64) \hspace{-0.05cm}\times\hspace{-0.05cm} 24 \hspace{-0.05cm}\times\hspace{-0.05cm} 48 \hspace{-0.05cm}\times\hspace{-0.05cm}
    16$,} & $M=1\hspace{0.05cm}(4)$ 
  \\ 
\end{tabular}
\end{footnotesize}
\label{CollParameterSet}
\end{ruledtabular}
\end{table}

\section{Effect of collisions in linear simulations}\label{SecCollLin} 
  
In this section, the effect of collisions on ITG eigenmodes is studied by performing a scan in collisionality. In the following subsection, the observed decrease in the linear growth rate with increasing collisionality is explained by using a velocity space analysis.

\subsection{Decrease in growth rate with increasing collisionality. Adiabatic-like electron response away from MRSs.}\label{SecCollLinGrowthRate}
\begin{figure}
	\centering
	\includegraphics[width=0.95\columnwidth] {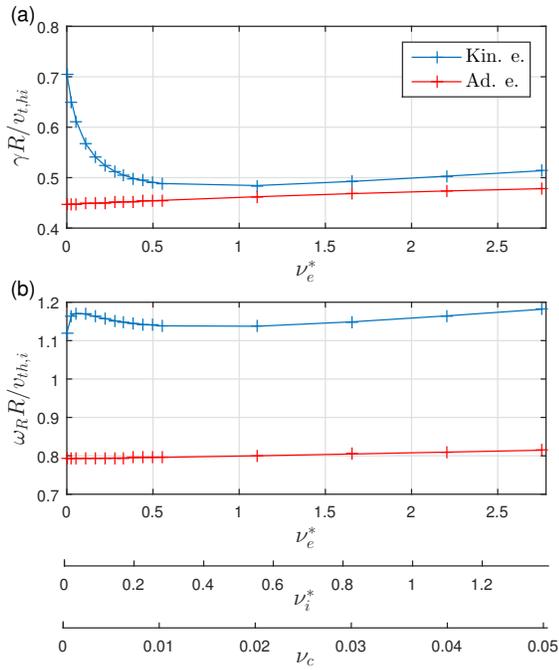}
	\caption{(a) Growth rate $\gamma$ and (b) real frequency $\omega_R$, in units of $v_{th,i}/R$, of ITG eigenmodes obtained for $k_y\rho_i=0.35$, plotted as a function of effective electron collisionality $\nu^*_e$ in linear simulations with either adiabatic (red) or kinetic (blue) electrons. Other simulation parameters are given in table ~\ref{CollParameterSet}. Axes with corresponding ion collisionality $\nu_i^*$ and GENE collisionality $\nu_c$ are also shown for comparison.}
	\label{gammaandomegavscoll}
	\end{figure}

In the collisionless case, non-adiabatic electron dynamics are known to strengthen the ITG instability drive \cite{Rewoldt1990,Mikkelsen2008,Dominski2015}. With the introduction of collisions, the instability drive is found to weaken, as has already been reported in Ref.~\onlinecite{Mikkelsen2008}. In Fig.~\ref{gammaandomegavscoll}, the growth rate of the ITG microinstability with kinetic electrons and $k_y\rho_i=0.35$ ($\sim$ most unstable) is shown to decrease with increasing collisionality for $\nu^*_e\lesssim 1$, and approaching the growth rate of the modes with adiabatic electrons for $\nu^*_e\sim 1$. The growth rate of the eigenmode with imposed adiabatic electron response presents a weak increase as collisionality is increased. In this subsection, a velocity space analysis of the electron distribution function is done to explain the weakening of the non-adiabatic electron instability drive with collisions. 

\begin{figure*}
	\centering
	\includegraphics[width=1.5\columnwidth] {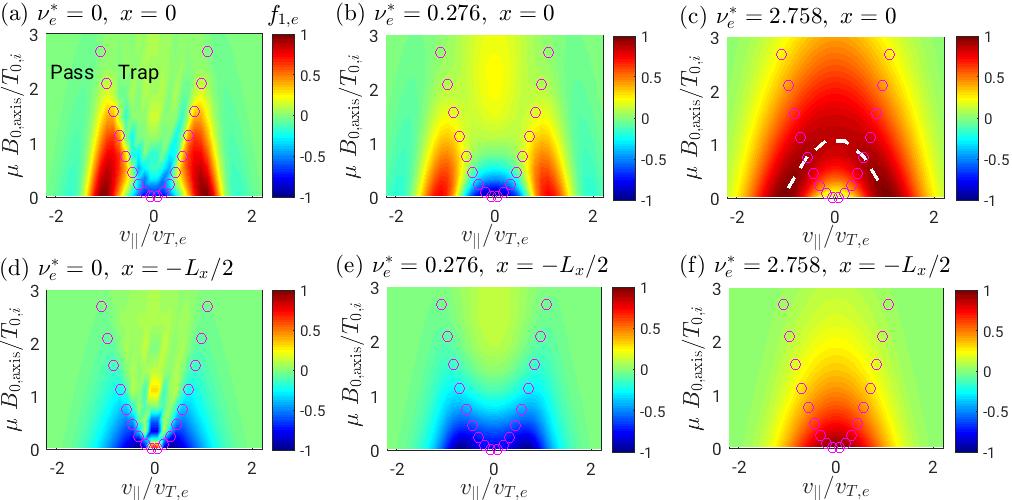}
	\caption{($v_{||},\mu$) velocity space colour plot of the real value of the perturbed electron distribution function $f_{1,e}$ of the eigenmodes in linear simulations with three different collisionalities $\nu_e^*=0,\ 0.276$ and $2.758$ (from left to right). The distribution function at the outboard mid-plane, \emph{i.e.} at $z=0$, is plotted for all the subfigures, with those at the top corresponding to radial location $x=0$, \emph{i.e.} at the corresponding MRS, and those at the bottom corresponding to $x=-L_x/2$, \emph{i.e.} mid-point between MRSs. Magenta circles indicate the boundary between trapped and passing regions. White dashed line indicates a constant energy curve $m_ev_\parallel^2/2+B_0\mu={\rm const.}$}
	\label{fonvpandmu}
	\end{figure*}

In Fig.~\ref{fonvpandmu}, the real value of the perturbed electron distribution function $f_{1,e}$ at the outboard midplane ($z=0$) for radial locations at ($x=0$) and away ($x=-L_x/2$) from MRS is plotted as a function of ($v_{\parallel},\mu$) for three cases, with collisionalities $\nu^*_e=0,0.276$ and $2.758$. In these plots, $v_\parallel$ is normalised by $v_{T,e}=\sqrt{2T_{0,e}/m_e}$ (not to be confused with $v_{th,e}=\sqrt{T_{0,e}/m_e}$). Magenta circles in the plots represent the boundary between the trapped and passing electron velocity space domains, defined by the relation $v_\parallel^2=(2/m_e)[B_{0}(z=-\pi)-B_0(z=0)]\mu$. 
Recall that in the flux-tube model, $B_0$ represents the equilibrium magnetic field strength on the flux-surface considered, and is independent of the radial coordinate $x$. Furthermore, $z=-\pi$ and $z=0$ denote the inboard midplane and outboard midplane where the magnetic field strength is maximum and minimum respectively, in the considered circular ad-hoc equilibrium~\cite{Lapillonne2009}. 
	
In the collisionless ITG case, using local dispersion relation, one can show that the trapped electrons are essentially passive and their response can be neglected \cite{Dominski2015}.  This is also illustrated in appendix~C of Ref.~\onlinecite{AjayCJPhD}, in particular in figure~C.4 where it is shown that the simplified slab like passing electron kinetic response \emph{SimPassKinE} result closely matches that including kinetic trapped electron response \emph{SimPassKinE + TrapKinE}. In Fig.~\ref{fonvpandmu}(a,d), while there is a small non-negligible contribution to the electron distribution function in the trapped region, most of the contribution is indeed localised in the passing electron velocity space, verifying that the trapped electrons are essentially passive in the collisionless case. Let us emphasize that, in linear simulations, in absence of collisions, passing and trapped electrons do no mix with each other. With collisions, \emph{i.e.} in Figs.~\ref{fonvpandmu}(b,c,e,f), the fluctuating part of the electron distribution function in the trapped region becomes comparable to the one in the passing region, especially at lower velocities where collisionality is stronger. This is a result of the collisional trapping-detrapping of electrons. 

	\begin{figure*}
	\centering
	\includegraphics[width=1.5\columnwidth] {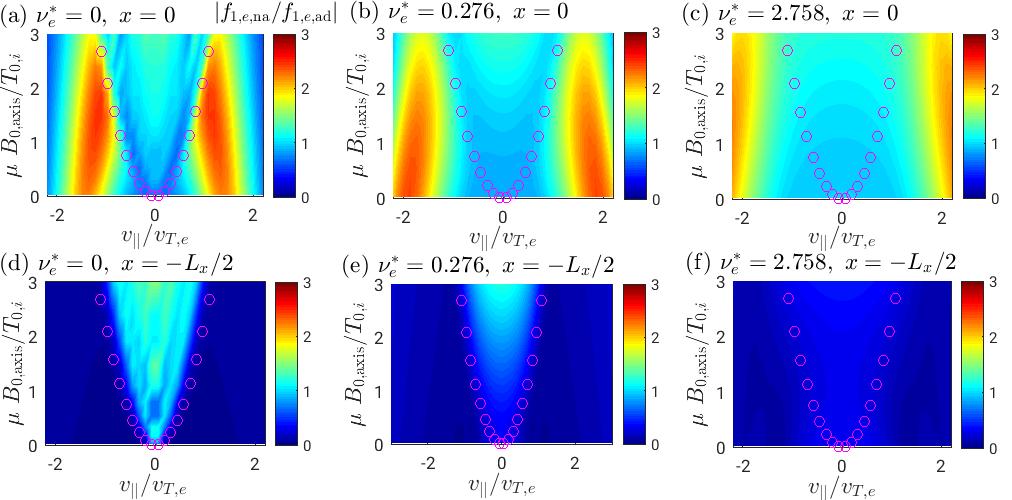}
	\caption{($v_{||},\mu$) velocity space contour plot of the non-adiabatic part of the perturbed electron distribution function $f_{1,e,\rm na}=f_{1,e}-f_{1,e,{\rm ad}}$ normalised by the adiabatic part $f_{1,e,\rm ad}$, in absolute value, \emph{i.e.} $|f_{1,e,\rm na}/f_{1,e,\rm ad}|$, of the eigenmodes in linear simulations presented in Figs.~\ref{PhiProfile_collcomp_rmr} and ~\ref{fonvpandmu}. Cases with three different collisionalities $\nu_e^*=0,\ 0.276$ and $2.758$ are shown, from left to right. The outboard mid-plane $z=0$ slice is plotted for all the subfigures, with those at the top corresponding to radial location $x=0$, \emph{i.e.} at the corresponding MRS, and those at the bottom corresponding to $x=-L_x/2$, \emph{i.e.} mid-point between MRSs. Magenta circles indicate the boundary between trapped and passing regions.}
	\label{fnonadonvpandmu}
	\end{figure*}

\emph{At MRS}, in the collisionless case, \emph{i.e.} in Fig.~\ref{fonvpandmu}(a), one can observe large relative amplitude of the perturbed electron distribution function at $v_{\parallel}\simeq \pm v_{T,e}$ in the passing electron domain of the velocity space. These structures are in fact characteristic of the non-adiabatic passing electron dynamics, as evident from the corresponding dominant structures in Fig.~\ref{fnonadonvpandmu}(a), where the non-adiabatic part of the perturbed electron distribution function $f_{1,e,\rm na}=f-f_{1,e,\rm ad}$ normalised by the adiabatic part $f_{1,e,\rm ad}=e\Phi f_{0,e}/T_0$ is plotted, where $f_{0,e}$ is the local Maxwellian. With collisions, the passing electrons get trapped and vice-versa, leading to a 'smearing' of these structures associated to non-adiabatic passing electrons in Fig.~\ref{fonvpandmu}(a) into the trapped electron domain of the velocity space as seen more so in Fig.~\ref{fonvpandmu}(c). In fact in Fig.~\ref{fonvpandmu}(c), one could clearly observe that the 'smeared' velocity space distribution of the electron distribution function follows the constant energy curve ($m_ev_\parallel^2+B_0\mu={\rm const.}$, denoted by the dashed white line), indicating that electron-ion pitch angle-scattering is the dominant collision mechanism at play.

	\begin{figure}
	\centering
	\includegraphics[width=1.25\columnwidth] {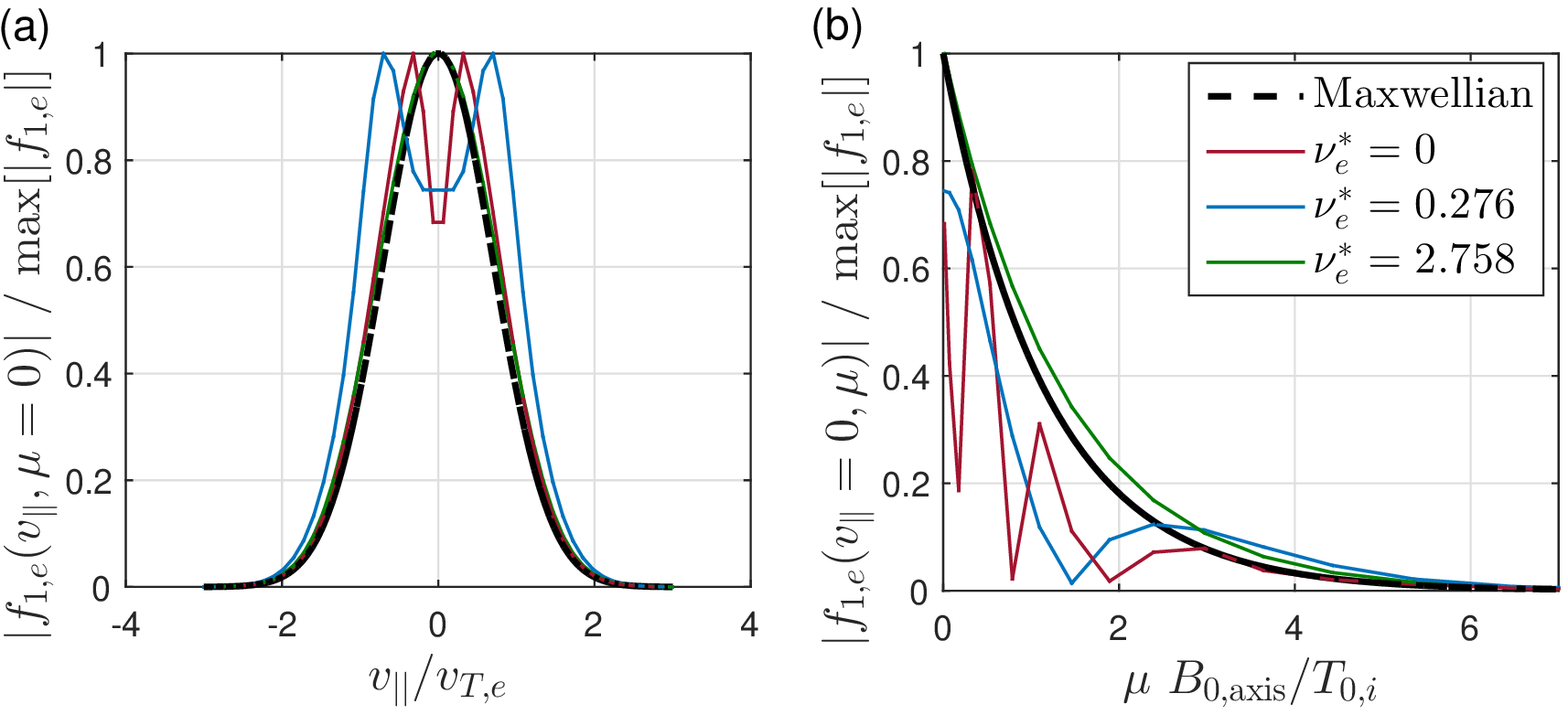}
	\vspace{-1.3cm}
	\caption{Velocity space dependence of the absolute value of the real value of the perturbed electron distribution function $f_{1,e}$, normalised by the maximum, for the eigenmodes in the same set of linear simulations presented in Fig.~\ref{fonvpandmu}, with collisionality $\nu_e^*=0$ (brown), $0.278$ (blue) and $2.758$ (green). In (a), the $v_\parallel$ profile is shown at $\mu=0$ and in (b), the $\mu$ velocity profile is shown at $v_{||}=0$, both at the outboard midplane $z=0$ and at the radial position $x=-L_x/2$ away from MRS. Dashed black line represent the normalised Maxwellian distribution.}
	\label{fvsvpandmu}
	\end{figure}

\emph{Away from MRS}, in the collisionless case, passing electrons behave adiabatically, as evident in Fig.~\ref{fnonadonvpandmu}(d). As collisionality increases, the collisional trapping-detrapping of electrons becomes more frequent and as one moves away from the banana regime there is less distinction between trapped and passing electrons. The trapped electrons as well therefore begin to respond adiabatically as can be seen in Fig.~\ref{fnonadonvpandmu}(e, f). 
The more adiabatic-like electron response away from MRS with increasing collisionality is also evident in Fig.~\ref{fvsvpandmu} which plots the $v_\parallel$ and $\mu$ profiles of the perturbed electron distribution function $f_{1,e}$ normalised by its maximum value, at $\mu=0$ and $v_\parallel=0$ respectively, at $x=-L_x/2$, at the outboard midplane. That is, with collisions, the distribution function becomes proportional to a Maxwellian, characteristic of adiabatic electron response. This is observed to a lesser extent at MRS as well. Naturally with increasing collisionality, as a greater fraction of electrons behave adiabatic-like, the growth rate decreases and approaches that for the fully adiabatic electron model.

\subsection{Increase in radial width of fine structures with increasing collisionality}\label{SecCollLinRadialwidth}

\begin{figure*}
	\centering
	\includegraphics[width=1.5\columnwidth] {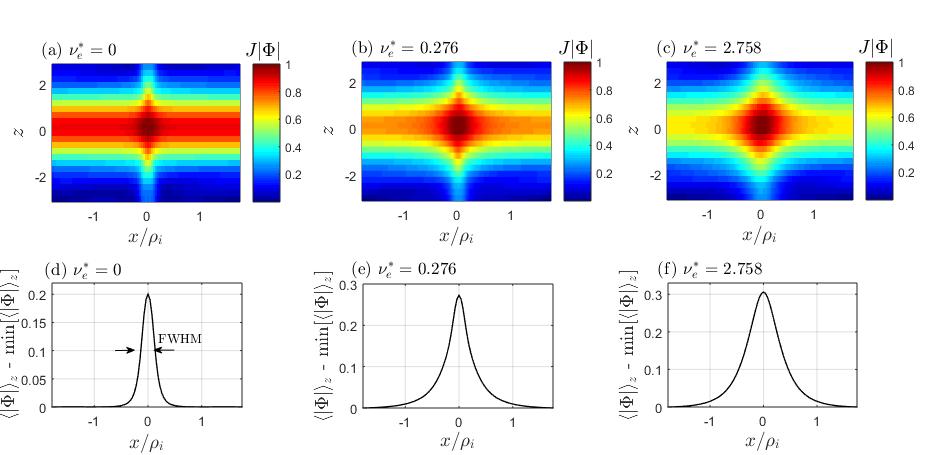}
	\caption{Top: $(x,z)$ colour plot of the absolute value of the electrostatic potential, (normalised by its maximum value and) weighted by the $(x,y,z)$ coordinate space Jacobian $J$, \emph{i.e.} $J |\Phi|$, for the eigenmodes in the same set of linear simulations in Fig.~\ref{fonvpandmu}, with collisionalities (a) $\nu_e^*=0$, (b) $\nu_e^*=0.276$ and (c) $\nu_e^*=2.758$. Bottom: The corresponding z-average of the absolute value of the electrostatic potential, subtracted by its minimum value, \emph{i.e.} $\langle|\Phi|\rangle_z$ - min[$\langle|\Phi|\rangle_z$], plotted as a function of $x$.}
	\label{PhiProfile_collcomp_rmr}
	\end{figure*}
	
\begin{figure}
	\centering
	\includegraphics[width=0.85\columnwidth] {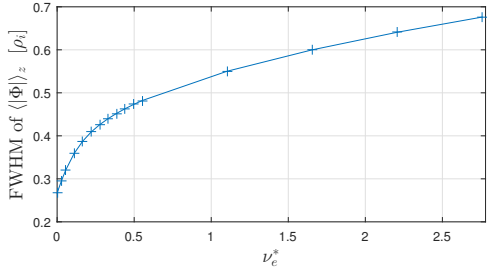}
	\caption{Full radial width at half maximum FWHM of the z-averaged electrostatic potential $\langle|\Phi|\rangle_z$ of the eigenmodes for the same set of linear simulations with kinetic electrons presented in Fig.~\ref{gammaandomegavscoll}.}
	\label{PhiWidth_FWHM_rmr}
	\end{figure}

With non-adiabatic passing electron dynamics, the ballooning structure of the ITG (and TEM) linear eigenmodes develop extended tails \cite{Hallatscheck2005}, corresponding to fine radial structures at associated MRSs \cite{Waltz2006,Chowdhury2008,Dominski2015,Dominski2017,AjayCJ2020}. Collisions are found to increase the radial width of these fine-structures. This is evident in Fig.~\ref{PhiProfile_collcomp_rmr}(a-c), which shows the $(x,z)$ dependence of the absolute value of the electrostatic potential $|\Phi|$ with different collisionalities, for the same eigenmodes considered in Figs.~\ref{fonvpandmu}. The corresponding $z$-averaged electrostatic potential  subtracted by its minimum value, \emph{i.e.} $\langle|\Phi|\rangle_z$ - min[$\langle|\Phi|\rangle_z$] is also plotted as a function of $x$ in Figs.~\ref{PhiProfile_collcomp_rmr}(d-f). To quantify the radial broadening of the fine-structures, the full width at half maximum FWHM of $\langle|\Phi|\rangle_z$ is plotted as a function of collisionality in Fig.~\ref{PhiWidth_FWHM_rmr}. In the following, this radial broadening of fine-structures is explained as the consequence of a decrease in the characteristic parallel length associated to the tail of the ballooning representation of the eigenmodes with increasing collisionality. 

\begin{figure}
	\centering
	\includegraphics[width=1\columnwidth] {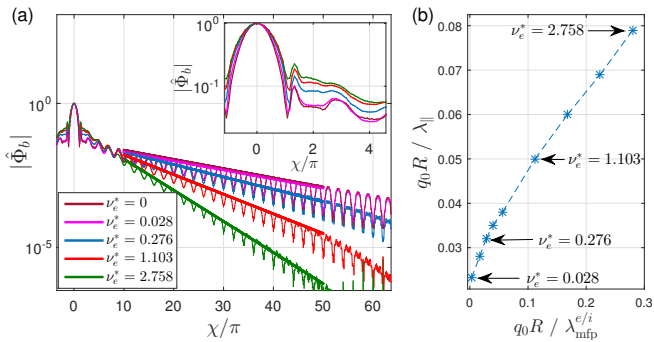}
	\caption{(a) Ballooning envelope $|\hat{\Phi}_b(\chi)|$ of the
    electrostatic potential $\Phi$	for the linear eigenmodes considered in
    Fig.~\ref{gammaandomegavscoll}, with collisionality $\nu_e^*=0$ (brown), 0.028(magenta), 0.276 (blue), 1.103 (red) and 2.758 (green). The $\nu_e^*=0$ and $0.028$ lines nearly overlap each other. Thick lines indicate the respective fits of the form $\mathcal{A} e^{-\chi/\Delta\chi}$. Zoom near $\chi=0$ is shown in the inset. (b) Inverse of characteristic parallel scale length $\lambda_\parallel=Rq_0\Delta\chi$ plotted as a function of the inverse of electron-ion mean free path $\lambda_{{\rm mfp},e}$, both normalised by $1/q_oR$.}
	\label{Ballooning_Philogvsz_collcomp}
	\end{figure}

In the ballooning representation, the radial Fourier ($k_x$) and the parallel (z) dependence of the linear mode profile is mapped to a purely parallel (ballooning space $\chi$) dependence [See Eq.~(\ref{ballooning defs.})]. In Fig.~\ref{Ballooning_Philogvsz_collcomp}(a), the ballooning representation of the electrostatic potential $|\hat{\Phi}_b|$ is plotted in log-lin scale as a function of the ballooning space angle $\chi$ for ITG eigenmodes with different collisionalities. While for $\chi<5$, the ballooning envelope is larger for higher collisionalities, for the significantly longer tail with $|\chi|\geq 5$, the ballooning envelope is smaller for higher collisionalities.

It can be seen that the tail of each of these ballooning structures presents essentially an exponential decay. 
A fit of the form $\mathcal{A}e^{-\chi/\Delta \chi}$ is therefore made, as indicated by the thicker lines in Fig.~\ref{Ballooning_Philogvsz_collcomp}(a); $\mathcal{A}$ is a constant and $\Delta \chi$ measures a characteristic parallel 'angular' extension of the mode in the extended ballooning space $\chi$. The angular width $\Delta \chi$ can be related to a characteristic length scale $\lambda_\parallel$, in the considered circular ad-hoc geometry, by the relation $\lambda_\parallel = Rq_0\Delta \chi$.

In Fig.~\ref{Ballooning_Philogvsz_collcomp}(b), the characteristic parallel length $\lambda_\parallel$ is shown to scale nearly linearly with the electron-ion mean free path $\lambda_{{\rm mfp}}^{e/i}=v_{th,e}/\nu_{ei}$, where $\nu_{ei}=\nu_{ei}(v_{T,e})$ is the thermal electron-ion collision frequency [relation between $\nu_{ei}$ and $\nu_e^*$ can be obtained from Eqs. (\ref{nueidef}) and (\ref{nustardef})]. That is, the electron-ion mean free path $\lambda_{{\rm mfp}}^{e/i}$ (or collisionality $\nu_e^*\propto 1/\lambda_{{\rm mfp}}^{e/i}$) sets the characteristic parallel length $\lambda_\parallel$ of the eigenmodes. 

To summarize, increasing collisionality leads to an increasing exponential decay rate $1/\Delta\chi$ ($\propto 1/\lambda_\parallel$) of the tail of the ballooning envelope. Given that a narrower tail in ballooning space is associated to broader radial fine-structures, one can therefore see how collisions lead to broadening of the fine-structures in real space.

\section{Effect of collisions in nonlinear simulations}\label{SecCollNonlin}
In this section, the effects of collisions in nonlinear turbulence simulations are studied, in particular, on the fine-structures associated to non-adiabatic passing electrons and the self-interaction mechanism. Towards this goal, a scan in collisionality $\nu_e^*\in\{0,0.028,0.276,2.758\}$ is performed, with the physical parameters as given in table~\ref{CollParameterSet}, simulating the same ITG dominant conditions as considered for the linear study. In the same table, numerical parameters wherever different from that in linear simulations have been given within parenthesis. The results are presented in the following four subsections.

\subsection{Effect of collisions on heat flux}\label{SecCollNLflux}

In this subsection, the dependence of ion heat flux on collisionality in nonlinear simulations with \emph{kinetic} and \emph{adiabatic} electron response is discussed. 

\begin{figure}
	\centering
	\includegraphics[width=0.8\columnwidth] {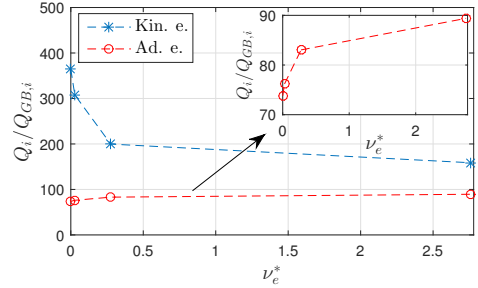}
	\caption{Time-averaged gyro-Bohm normalised ion heat flux $Q_i/Q_{GB,i}$ as a function of collisionality $\nu_e^*$ in simulations with kinetic (blue asterisks) and adiabatic (red circles) electrons.}
	\label{Qivscoll}
	\end{figure}

In simulations with \emph{kinetic} electrons, the gyro-Bohm normalised ion heat flux is found to decrease with increasing collisionality, as shown in Fig.~\ref{Qivscoll}. In Ref.~\onlinecite{Mikkelsen2008}, this drop in heat flux with collisionality is attributed to the corresponding reduction in the linear growth rates (discussed in section~\ref{SecCollLinGrowthRate}, see Fig.~\ref{gammaandomegavscoll}). This statement is further justified by  an analysis based on quasilinear estimate of flux levels and another based on the shearing rate associated with $E\times B$ zonal flows in subsections~\ref{SecQLan} and \ref{SecZonFloAn} respectively. 

\subsubsection{Quasilinear analysis}\label{SecQLan}
To verify that the decrease in linear growth rates with collisionality leads to the decrease in nonlinear heat and particle fluxes in kinetic electron simulations, a quasilinear estimate of the flux levels is performed. Following the model in Refs.~\onlinecite{Fable2009,Lapillonne2011,Mariani2018}, one has the quasilinear ion heat flux 
\begin{align}
Q^{QL}_i &= A_0\sum_{k_y} \hat{Q}^{QL}_i(k_y), \ \ \ {\rm where} \\
\hat{Q}^{QL}_i(k_y) &= w^{QL}(k_y)Q^L_{\rm norm}(k_y),
\end{align}
and $A_0$ is a constant. $w^{QL}(k_y)$ are quasilinear weights modelling the relative saturation levels of the nonlinear electrostatic potential for each $k_y$. The following form for these weights are considered
\begin{align}
w^{QL}(k_y)=
\left( \frac{\gamma(k_y)}{\langle k_\perp^2\rangle (k_y)}
\right)^\xi,
\end{align}
where $\gamma$ is the growth rate of the eigenmode for each $k_y$ and 
\begin{align}
\langle k_\perp^2\rangle (k_y) 
= \frac{\sum_{k_x}\int k_\perp^2(k_x,k_y,z)|\hat{\Phi}(k_x,k_y,z)|^2 J(z) dz}{\sum_{k_x}\int |\hat{\Phi}(k_x,k_y,z)|^2 J(z) dz}
\label{kpfs}
\end{align}
is the squared perpendicular wavenumber $k_\perp^2(k_x,k_y,z)=g^{xx}(z)k_x^2 + 2g^{xy}(z)k_xk_y + g^{yy}(z)k_y^2$ weighted-averaged over the mode structure envelope $|\hat{\Phi}|^2$. $g^{\mu\nu}=\nabla\mu\cdot\nabla\nu$ for $\nu,\mu=x,y,z$ are the metric coefficients. $\xi$ is a fit parameter chosen after matching the resulting quasilinear $k_y$ spectra with the corresponding nonlinear spectra. The typical values of $\xi$ considered are $1,\ 2$ and $3$ \cite{Jenko2005,Casati2009,Fable2009,Lapillonne2011,Mariani2018}.

$Q^L_{\rm norm}(k_y)$ is the normalised `linear heat flux' computed from the linear eigenmodes as
\begin{align}
Q^L_{\rm norm}(k_y) = \frac{1}{|\hat{\Phi}(k_x=0,k_y,z=0)|^2}Q^L(k_y),\ \ \ {\rm with} \\
Q^{L}(k_y) = \left\langle
\frac{1}{\mathcal{C}}\sum_{k_x} 2\ {\rm Re}
\left[ ik_y\hat{\Phi}^*
\int \frac{1}{2}m v^2 \hat{\delta f} d^3v
\right]
\right\rangle_z,
\label{QLky}
\end{align}
where $\langle\cdot\rangle_z=\int\cdot J(z)dz/\int J(z)dz$, $\mathcal{C}=B_0/\sqrt{g^{xx}g^{xy}-(g^{xy})^2}$, ${\rm Re}$ indicates the real part of a function, $m$ is the mass of ions and  $\hat{\delta f}$ is the fluctuating part of the particle distribution function. In Eqs.~(\ref{kpfs}) and (\ref{QLky}), the summation over $k_x$ involves the linearly coupled radial modes for a given $k_y$ and can be explicitly written as $\sum_{kx}=\sum_{p=-N_{kx}^{QL}}^{N^{QL}_{kx}}$ with $k_x=p 2\pi k_y\hat{s}$. 

\begin{figure}
	\centering
	\includegraphics[width=1.05\columnwidth] {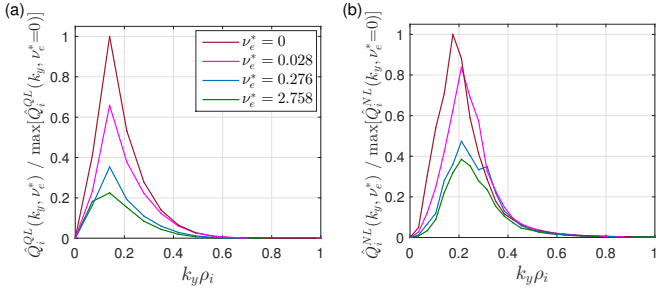}
	\caption{(a) $k_y$ spectra $\hat{Q}^{QL}_i(k_y,\nu_e^*)$ of the ion heat flux quasilinear estimate with $N^{QL}_{kx}=1$ and $\xi=2$, for four different values of collisionality $\nu_e^*=0$ (brown), $0.028$ (magenta), $0.276$ (blue) and $2.758$ (green), all normalised by the maximum of the collisionless case. (b) $k_y$ spectra $\hat{Q}^{NL}_i(k_y,\nu_e^*)$ of saturated nonlinear ion heat flux for the same four different values of collisionality considered, all normalised by the maximum of the collisionless case.}
	\label{QLest}
	\end{figure}

Following Ref.~\onlinecite{Mariani2018}, three different values of $N^{QL}_{kx}=0,1$ and $N_{kx}$ are considered here. Three different values of $\xi=1,\ 2$ and $3$ are also considered. The resulting 9 sets of quasilinear ion heat flux $k_y$ spectra $\hat{Q}^{QL}_i(k_y)$ for the collisionless case  were then compared with the corresponding saturated nonlinear ion heat flux spectra. From this analysis it was found that the $N^{QL}_{kx}=1$ and $\xi=2$ gave the best fit. This is shown in Fig.~\ref{QLest}(a) and (b) where the resulting quasilinear and saturated nonlinear $k_y$ ion heat flux spectra respectively for the collisionless case, both normalised by their maximum, are plotted in brown. The plots for the three finite collisionalities $\nu_e^*=0.028,\ 0.276$ and $2.758$ are also shown, normalised by the maximum of their respective collisionless cases.

The relative decrease in the total quasilinear ion heat flux $Q^{QL}_i$, assuming $A_0$ to remain the same, are $30.5\%,\ 62.6\%$ and $70.9\%$ for $\nu_e^*=0.028,\ 0.276$ and $2.758$ respectively, with respect to $\nu_e^*=0$. Note that the overall fit parameter $A_0$ that should account for the final balance between the drive and saturation mechanisms is in general itself a function of collisionality. We will however assume here that $A_0$ remains approximately constant over the considered collisionality scan. The decrease with respect to $\nu_e^*=0$ in total saturated nonlinear ion heat flux for the runs with kinetic electrons are $15.1\%,\ 45.2\%$ and $54.8\%$ for $\nu_e^*=0.028,\ 0.276$ and $2.758$ respectively. There is thus a fairly good semi-quantitative agreement between the relative decrease in the quasilinear and nonlinear flux levels, at least approximately correct to a factor of 2, thereby verifying the assumption of constant $A_0$ to a good degree and also providing a validation for the argument that  decrease in linear growth rates with collisionality is the main cause for the decrease in nonlinear flux levels in kinetic electron simulations.


Note that quasilinear estimates sometimes fail to capture nonlinear saturation mechanism such as for example the zonal flow saturation mechanism in the Dimits shifted regime. This is based on the fact that in the Dimits shifted regime where the nonlinear flux levels are zero, the quasilinear estimate which computes the fluxes primarily based on the growth rates of linear eigenmodes still predicts non-zero flux levels. Hence, the quasilinear analysis presented above might fail to capture the effects of zonal flow saturation. Therefore, complementing this quasilinear analysis, in the following, an analysis based on shearing rate associated with zonal flows is performed to further verify that the decrease in nonlinear heat flux with increasing collisionality is indeed a result of the decrease in the growth rate of corresponding linear eigenmodes.

\subsubsection{Zonal flow saturation analysis}\label{SecZonFloAn}
Shearing of turbulent eddies by zonal flows is a primary mechanism by which ITG driven turbulence saturates \cite{Biglari1990,Rosenbluth1998,Lin1998,Waltz1994}. 

For a quantitative analysis, the effective shearing rate $\omega_{\rm eff}$ defined in Ref.~\onlinecite{AjayCJ2020} is considered. This rate is estimated as follows. One first
defines the zonal $E\times B$ shearing rate experienced by the ions, which
are the dominant instability drivers in the case of ITG
turbulence considered here:
  \begin{equation}
  \label{Eq_wExBion}
  \omega_{E\times B,{\rm ion}}(x, t) 
  =
  \frac{1}{B_0}\,
  \frac{\partial^2\langle\bar{\Phi}\rangle_{y,z}}{\partial x^2},
  \end{equation}
where the flux-surface average $\langle\bar{\Phi}\rangle_{y,z}$ provides the
zonal component of $\bar{\Phi}$ and involves both an average over $y$,
$\langle\cdot\rangle_y = (1/L_y)\int_0^{L_y} \cdot\,dy$, and an average over
$z$, \\
$\langle\cdot\rangle_z = \int_{-\pi}^{+\pi}\cdot\,J
dz/\int_{-\pi}^{+\pi} Jdz$, with $J$ being the $(x,y,z)$ coordinate space Jacobian. $\bar{\Phi}$ is the scalar potential
gyro-averaged over the Maxwellian ion background velocity distribution, \emph{i.e.} $\bar{\Phi}=\mathcal{G}_i\Phi$, where the gyro-averaging operator $\mathcal{G}_i$ itself involves averaging over the Maxwellian background distribution, given in Fourier space by
  \begin{align*}
  \hat{\mathcal{G}}= & \int_0^\infty v_\perp d v_\perp \ 
  e^{
  -\frac{1}{2}
  \left(  \frac{v_\perp}{v_{th,i}}\right)^2
  }
  \ J_{0}
  \left(\frac{k_\perp v_\perp}{\Omega_i}
  \right)
  = 
  e^{
  -\frac{(k_\perp \rho_i)^2}{4}
  }
  \end{align*}
Since the effective electric field felt by the ions is a gyro-averaged one, it is justified to consider the gyro-average of the scalar potential. It also eliminates the observed non-vanishing tail in the $k_x$-spectra of $\omega_{E\times B}$ for $|k_x\rho_i\rightarrow \infty|$ when a non-gyroaveraged scalar potential is considered.
  
The shearing rate $\omega_{E\times B,{\rm ion}}$ is then furthermore averaged
over a small time window of width $\tau$, given that fluctuations that are
very short-lived in time do not contribute effectively towards the zonal
flow saturation mechanism \cite{Hahm1999}, thus providing the effective
shearing rate:
  \begin{equation}
  \label{effsheardef}
  \omega_{\rm eff}(x, t) 
  = 
  \frac{1}{\tau}
  \int_{t-\tau/2}^{t+\tau/2} \hspace{-0.5cm}
  \omega_{E\times B,{\rm ion}}(x, t')\,dt'.
  \end{equation}
Here, $\tau=1/\gamma_{\max}$ is considered, where $\gamma_{\max}$ is the
growth rate of the most unstable linear mode.

Three different estimates of radial and time averages of $\omega_{\rm eff}$ are considered: 
(a) time and system average of total shearing rate ${\rm RMS}_{x,t}(\omega_{\rm eff}) = \left(\, \langle\, \omega_{\rm eff}^2\,\rangle_{x,t} \,\right)^{1/2}$, 
(b) contribution from the stationary components, ${\rm RMS}_{x}(\langle\omega_{\rm eff}\rangle_t) = \left[\,\langle\, \left(\, \langle\omega_{\rm eff}\rangle_t \,\right)^2\,\rangle_{x} \,\right]^{1/2}$ and 
(c) contribution from fluctuating components ${\rm SD}_{x,t}(\omega_{\rm eff}) = \left[\, \langle\, \left(\omega_{\rm eff} - \langle\omega_{\rm eff}\rangle_t \right)^2\,\rangle_{x,t} \,\right]^{1/2}$, all 
normalised by their corresponding maximum linear growth rates. Note ${\rm RMS}_{x,t}^2(\omega_{\rm eff}) = {\rm RMS}_x^2(\langle \omega_{\rm eff}\rangle_t) + {\rm SD}_{x,t}^2(\omega_{\rm eff})$.

\begin{figure*}
	\centering
	\includegraphics[width=1.85\columnwidth] {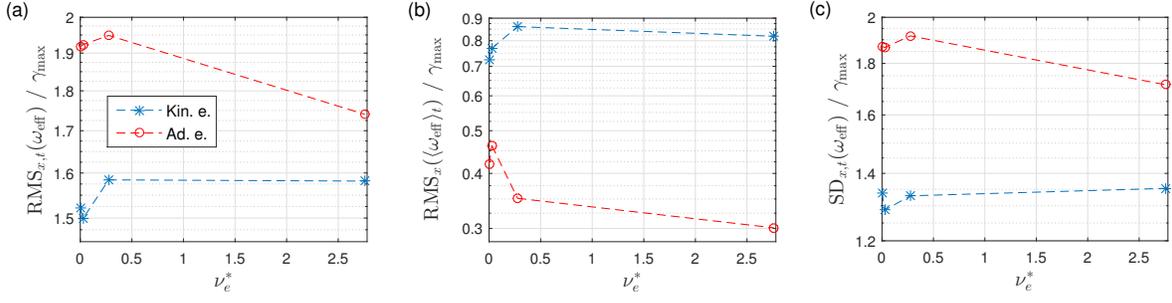}
	\caption{Effective shearing rate $\omega_{\rm eff}$ associated to the zonal
    $E\times B$ flows, normalised to corresponding maximum linear growth rate
    $\gamma_{\max}$, as a function of collisionality $\nu_e^*$. Blue asterisks denote
    kinetic electron simulations and red circles denote adiabatic electron
    simulations. (a) Time and system average of total shearing rate estimated with ${\rm
      RMS}_{x,t}(\omega_{\rm eff}) = \left(\, \langle\, \omega_{\rm eff}^2
    \,\rangle_{x,t} \,\right)^{1/2}$. (b) Contribution from the stationary
    component, ${\rm RMS}_{x}(\langle\omega_{\rm eff}\rangle_t) = \left[\,
      \langle\, \left(\, \langle\omega_{\rm eff}\rangle_t \,\right)^2
      \,\rangle_{x} \,\right]^{1/2}$. (c) Contribution from fluctuation
    component, ${\rm SD}_{x,t}(\omega_{\rm eff}) = \left[\, \langle\, \left(
      \omega_{\rm eff} - \langle\omega_{\rm eff}\rangle_t \right)^2
      \,\rangle_{x,t} \,\right]^{1/2}$.}
	\label{wvscoll}
	\end{figure*}

These estimates of the effective shearing rate for the considered nonlinear simulation scan over collisionality are shown in Figs.~\ref{wvscoll}(a-c). The plot of the shearing rate contribution from stationary structures for the case with kinetic electrons in Fig.~\ref{wvscoll}(b) shows an increase of $19\%$ as collisionality increases from $\nu_e^*=0$ to 0.276, which then drops to an increase of only $13\%$ between $\nu_e^*=0$ and 2.758. The corresponding ion heat flux plot in Fig.~\ref{Qivscoll}, on the other hand shows a monotonic decrease over the full considered range of collisionality from $\nu_e^*=0$ to 2.758. The fluctuating component of zonal flows estimated with ${\rm SD}_{x,t}(\omega_{\rm eff})$, which also play an important role in the saturation mechanism, as explained in detail in section 4 of Ref.~\onlinecite{AjayCJ2020}, is found to show a negligible maximum change of only $6\%$ in Fig.~\ref{wvscoll}(c). The total shearing rate estimate in Fig.~\ref{wvscoll}(a) also shows only a $5\%$ maximum change over the considered range of collisionalities. These results suggest that, for the considered simulations, turbulence saturation via zonal flows is less likely to be the primary factor leading to the decrease in heat flux with increasing collisionality. The decrease in linear drive of turbulence with increasing collisionality is therefore most likely to be the reason for the observed decrease in turbulent fluxes, which further validates the quasi-linear model applied in the previous section for interpreting the nonlinear simulations.

\subsection{Effect of collisions on radial width of fine-structures}\label{SecCollNonlinFSWidth}
The increase in the radial width of fine-structures on linear kinetic electron ITG eigenmodes with increasing collisionality have already been discussed in section~\ref{SecCollLinRadialwidth}. In this subsection, the effect of collisions on the width of these structures in the turbulent steady state of nonlinear simulations is explored.

Recalling the definition of ballooning representation in Eq.~(\ref{ballooning defs.}), one notes that a broader tail of the ballooning envelope of an eigenmode reflects a radially narrower fine structure in real space. It is therefore also possible to study the effect of collisionality on the width of fine-structures associated to an eigenmode in nonlinear simulations by comparing its ballooning representations across simulations with different collisionalities.

\begin{figure}
	\begin{minipage}{0.5\textwidth}
	\centering
	\includegraphics[width=0.95\columnwidth] {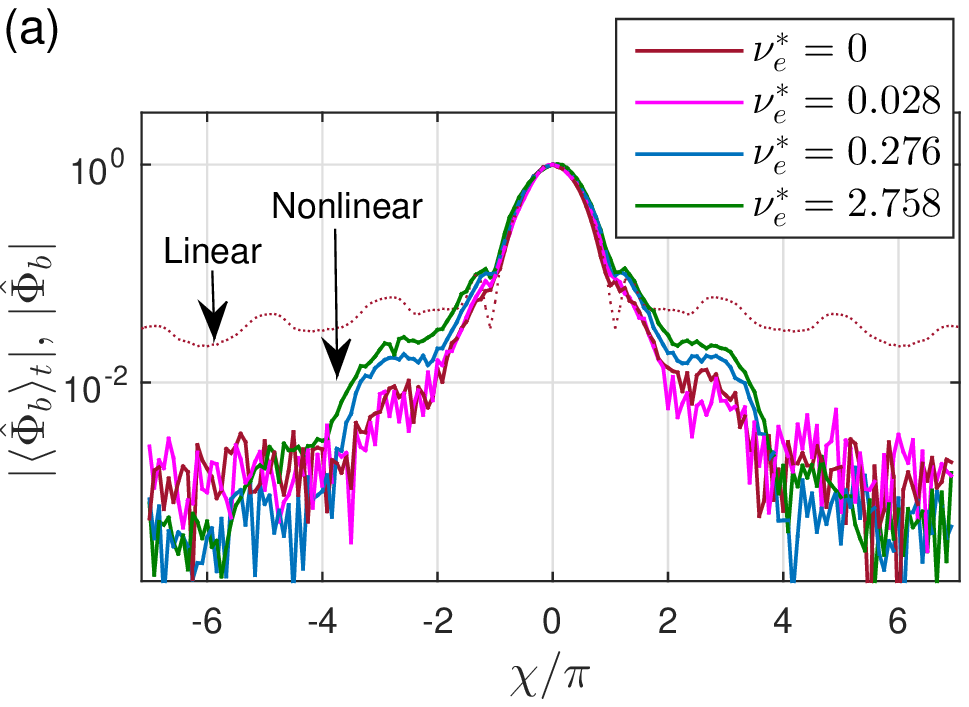}
	\end{minipage}
	\\
	\begin{minipage}{0.5\textwidth}
	\includegraphics[width=0.75\columnwidth] {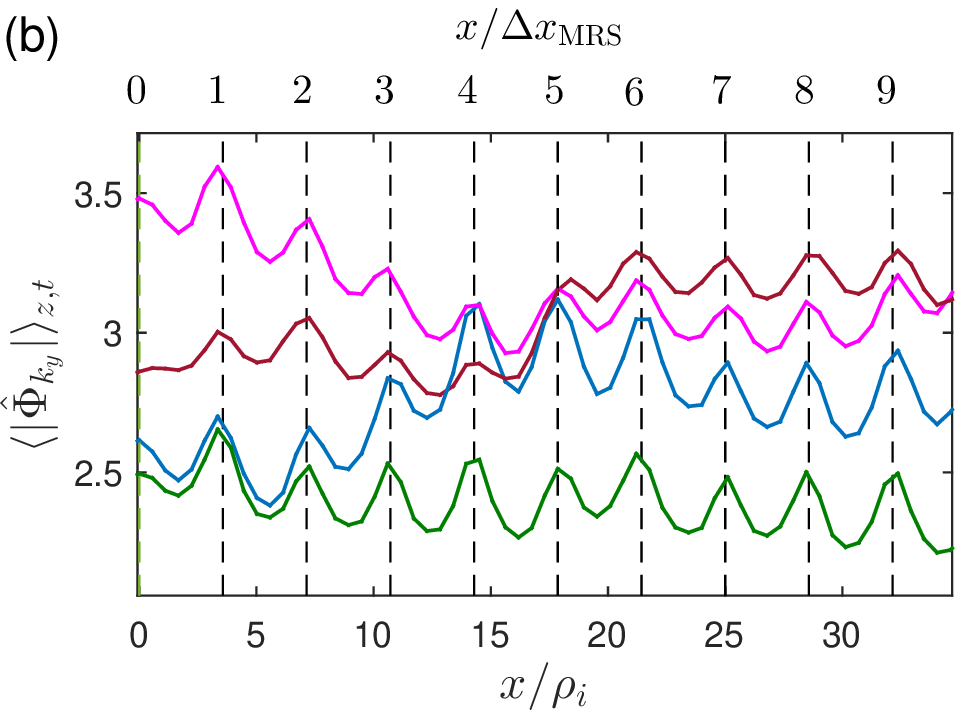}
	\end{minipage}
	\caption{(a) Absolute value of the time-averaged ballooning
structure of the electrostatic potential normalised by its value at $\chi=0$, \emph{i.e.}
$|\langle\hat{\Phi}_{b}(\chi,t)/\hat{\Phi}_{b}(\chi_0=0,t)\rangle_{t}|$, 
for $k_{x0}=0$ and $k_y\rho_i=0.35$ in turbulence simulations with collisionalities $\nu_e^*=0$ (brown), $0.028$ (magenta), $0.276$ (blue) and $2.758$ (green). (b) Time and $z$-averaged electrostatic potential $\langle|\hat{\Phi}_{k_y}|\rangle_{z,t}$ plotted as a function of the radial coordinate $x$ (only a sub-interval of the full radial box length $L_x=142.9\rho_i$ is shown), for the same set of simulations as in sub-plot (a), and with the same colour labels.}
	\label{Ballooning_nltavg}
	\end{figure}

In Fig.~\ref{Ballooning_nltavg}(a), the absolute value of the time-averaged ballooning representation of the electrostatic potential for $k_{x0}=0$ and $k_y\rho_i=0.35$ (same as in Fig.~\ref{Ballooning_Philogvsz_collcomp}), normalised by its value at $\chi=0$, \emph{i.e.} $|\langle\hat{\Phi}_b(\chi,t)/\hat{\Phi}_b(\chi_0=0,t)\rangle_t|$, is plotted for the four different values of collisionality considered. Contrary to the linear result, a narrowing of the extended ballooning tail, corresponding to a radial \emph{broadening} in real space, is not observed with increasing collisionality. In fact a slight \emph{narrowing} of the radial width is observed (visible in Fig.~\ref{Ballooning_nltavg}(b)). This different dependence of radial width on collisionality between linear and nonlinear results is a consequence of the linear coupling between $k_x=k_{x0}+p2\pi\hat{s}k_y, p\in\mathbb{Z}$, in an eigenmode being significantly disrupted by nonlinear couplings in the turbulent state for $|p|\geq 2$. 

Furthermore, recall that in the case of linear eigenmodes discussed in section~\ref{SecCollLinRadialwidth} with the help of Fig.~\ref{Ballooning_Philogvsz_collcomp}(a), lower relative values of the ballooning envelope are observed with increasing collisionality for $|p|\geq 2$, or more precisely for $|\chi|\geq 5\pi$, while higher values are observed for $|\chi|< 5\pi$. The latter is similar to that observed in the corresponding time-averaged ballooning envelope in nonlinear simulations in Fig.~\ref{Ballooning_nltavg}(a), and consistent with the radial broadening of the time-averaged fine-structures in real space.

%


To further study the nonlinear modification of an eigenmode, one can measure how well the linear phase difference along its ballooning structure is retained in a nonlinear simulation. The relative phase $\delta\phi_{\rm nl}$ along the ballooning structure in nonlinear simulations, defined as 
$\delta\phi_{\rm nl}(\chi,t)=\phi[\hat{\Phi}_{b,{\rm nl}}(\chi,t)/\hat{\Phi}_{b,{\rm nl}}(\chi_0=0,t)]$, is therefore compared to its linear value
$\delta\phi_{\rm lin}(\chi)=\phi[\hat{\Phi}_{b,{\rm lin}}(\chi)/\hat{\Phi}_{b,{\rm lin}}(\chi_0=0)]$;
Here, $\phi[A]$=arg($A)$ stands for the phase or argument of the complex number $A$.

\begin{figure*}
	\centering
	\includegraphics[width=1.8\columnwidth] {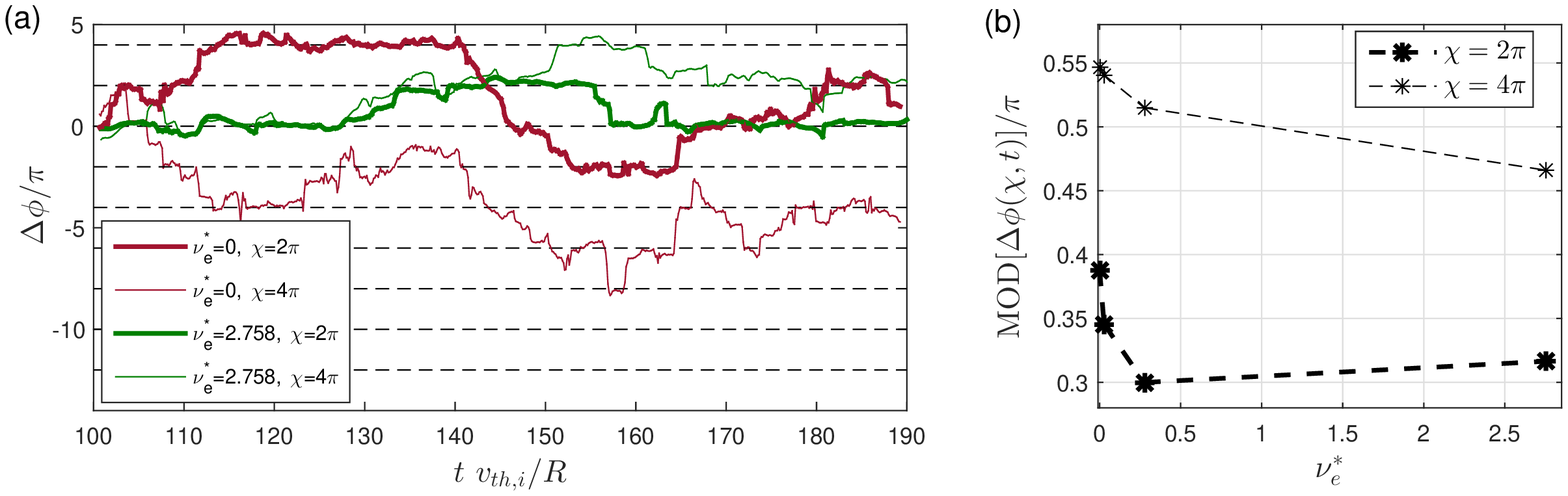}
	\caption{Phase difference $\Delta\phi(\chi,t)=(\phi[(\hat{\Phi}_{b,{\rm nl}}(\chi,t)/\hat{\Phi}_{b,{\rm nl}}(\chi_0=0,t))(\hat{\Phi}_{b,{\rm lin}}(\chi_0=0)/\hat{\Phi}_{b,{\rm lin}}(\chi))])$ plotted as a function of time. Here, $\hat{\Phi}_{b,{\rm nl}}$ and $\hat{\Phi}_{b,{\rm lin}}$ denote the ballooning representation of the electrostatic potential in non-linear and linear simulations respectively, for the eigenmode with $k_{x,0}=0$ and $k_{y,\rho_i}=0.35$.  Brown and green colours represent simulations with collisionality $\nu_e^*=0$ and $2.758$ respectively, while thick and thin lines represent $\chi=2\pi$ and $4\pi$ respectively. (b) MOD$(\Delta\phi(\chi,t))$ plotted as a function of collisionality $\nu_e^*$. Thick and thin lines represent $\chi=2\pi$ and $4\pi$ respectively.}
	\label{Phasediffvst_collcomp}
	\end{figure*}

In Fig.~\ref{Phasediffvst_collcomp}(a), the quantity $\Delta\phi(\chi,t)=\delta\phi_{\rm nl}(\chi,t)-\delta\phi_{\rm lin}(\chi)$ measuring the deviation of the relative phase, at the extended ballooning space coordinate $\chi$, between the nonlinear and linear simulations, is plotted as a function of time for $\chi=2\pi$ and $4\pi$. The eigenmodes with $k_{x0}=0$ and $k_y\rho_i=0.35$ is considered for two nonlinear simulations corresponding to the two extreme values of collisionalities considered here, \emph{i.e.} $\nu_e^*=0$ and $\nu_e^*=2.758$. In this figure, one can observe that the relative phase along the ballooning structure remains nearer to its linear value in the case with $\nu_e^*=2.758$ (as indicated by the corresponding plots more closely adhering to one of the horizontal dashed lines representing phase differences that are multiples of $2\pi$), than in the case with no collisions. 

To quantitatively measure the extent to which the linear phase difference along the ballooning structure is retained in nonlinear simulations, one can use the quantity MOD$(\Delta\phi(\chi,t))$, defined in Eq.~(\ref{defMOD}).
\begin{equation}
\label{defMOD} 
{\rm MOD}(A)=(\langle|{\rm mod}_{2\pi}(A)|^2\rangle_t)^{1/2},
\end{equation}
where mod$_{2\pi}(A) \equiv A-2\pi \times {\rm round}(A/2\pi)$, $A\in\mathbb{R}$, and the function round provides the nearest integer.
Note that, the smaller the value of MOD$(\Delta\phi(\chi,t))$, the more strongly the relative phase difference is fixed by the linear couplings, while for uniform random values of $\Delta\Phi$ between $-\pi$ and $\pi$, one obtains MOD$(\Delta\phi(\chi,t))=0.58\pi$. In Fig.~\ref{Phasediffvst_collcomp}(b), MOD$(\Delta\phi(\chi,t))$ is plotted as a function of collisionality $\nu_e^*$, for $\chi=2\pi$ and $4\pi$. From this figure, in general (with the exception of the $\nu_e^*=2.758$ data point for the case with $\chi=2\pi$), one can conclude that the linear relative phase difference along the ballooning structure of an eigenmode is more closely maintained in turbulence simulations with larger collisionalities. 

Figs.~\ref{Phasediffvst_collcomp} (a) and (b) show that collisions affect the nonlinear modification of the eigenmodes, in particular leading to a decrease in the width of the fine-structures with increasing collisionality.
This can be seen explicitly in Fig.~\ref{Ballooning_nltavg}(b), where the time and $z$ averaged absolute value of the electrostatic potential $\langle|\hat{\Phi}_{k_y}|\rangle_{z,t}$ for $k_y\rho_i=0.35$ is plotted as a function of $x$; $\hat{\Phi}_{k_y}$ is defined as per the relation
    \begin{equation}
	\label{Phikydef}
        \Phi(x,y,z)=\sum_{k_y}\hat{\Phi}_{k_y}(x,z)e^{ik_yy}.
    \end{equation}
Only a part of the radial domain is shown in this figure for better visualization. Each of the peaks is located at the radial positions of the corresponding MRSs. For the collisionless case, these fine-structures can be seen to be flatter, which then become more peaked as collisionality increases, indicating a (slight) narrowing of the fine-structures with collisions.

\begin{figure*}
	\centering
	\hspace{-0.5cm}
	\includegraphics[trim= 0cm 0cm 7cm 0cm, clip=true, width=1.75\columnwidth]{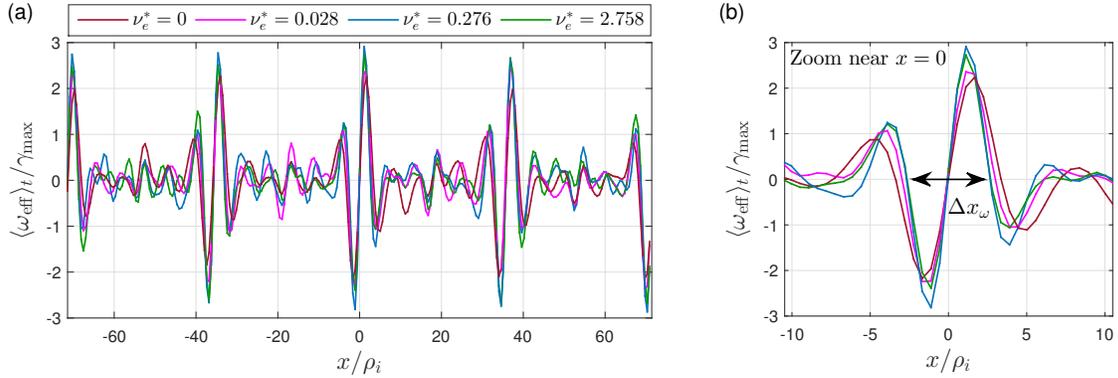}
	\vspace{-1.25cm}
	\caption{(a) Radial profile of the time-averaged effective shearing rate $\langle\omega_{\rm eff}\rangle_t$ normalised by the respective maximum linear growth rates $\gamma_{\rm max}$
	in simulations with collisionalities $\nu_e^*=0$ (brown), $0.028$ (magenta), $0.276$ (blue) and $2.758$ (green). (b) Zoom of the same plot near $x=0$.}
	\label{wvsx_collcomp}
	\end{figure*}

A slight decrease in the width of fine-structures with increasing collisionality can be further observed on the shearing rate associated to zonal flows, as seen in Fig.~\ref{wvsx_collcomp}, where the effective shearing rate $\omega_{\rm eff}$ is plotted as a function of the radial coordinate x for different collisionalities. 

\subsection{Effect of collisions on self-interaction mechanism}\label{SecCollNLSI}

Self-interaction is the process by which each microturbulence eigenmode interacts non-linearly with itself to produce a Reynolds stress contribution to zonal flow drive. While a detailed explanation of the self-interaction mechanism can be found in Ref.~\onlinecite{AjayCJ2020}, a short description is provided in the following paragraph. 

One may recall the general form of an eigenmode in Eq.~(\ref{linear eigenmode, Fourier rep.}), in particular the linear coupling of $k_x=k_{x0} + p\ 2\pi k_y\hat{s}$ Fourier modes, with $p\in\mathbb{Z}$. In self-interaction, any two such Fourier modes $\hat{\Phi}_{k_{x0}  + p\,2\pi k_y\hat{s}, \,k_y}$ and $\hat{\Phi}_{k_{x0} + p''\,2\pi k_y\hat{s},\,k_y}$ composing the physical eigenmode will drive, via three Fourier mode coupling, the zonal mode $\hat{\Phi}_{p'\,2\pi k_y\hat{s},\,0}$ with $p'=p-p''$. Note that this drive of zonal modes is via the same quadratic non-linearity in the gyrokinetic equation related to the $E\times B$ drifts. Since the relative phases between the Fourier modes $\hat{\Phi}_{k_{x,0} + p\,2\pi k_y\hat{s},\,k_y}$ remains set to some extent by the linear coupling even during the non-linear turbulent evolution, the phases of the associated contributions to Reynolds stress driving the zonal modes are fixed. This translates in direct space to an essentially fixed periodic radial dependence (with period $\Delta x_{\rm MRS} = 1/k_y\hat{s}$ corresponding to the distance between MRSs, as seen in Fig.~\ref{d2RSsibyrms_collcomp}) of the contribution to Reynolds stress from a given $k_y$ eigenmode via this self-interaction mechanism. 

In this subsection, the effects of collisions on the self-interaction mechanism in nonlinear simulations are explored with the help of three diagnostics, namely the normalised self-interaction contribution to Reynolds stress presented in section~\ref{SecNormSIRS}, the bicoherence type analysis presented in section~\ref{SecBicoh} and the analysis based on the correlation between the different $k_y$ contributions to Reynolds stress presented in section~\ref{SecCorrRS}.


As will be seen in these following sections, it may \emph{a priori} appear that there is a contradiction of results from these diagnostics, with the first diagnostic indicating that the relative contribution to Reynolds stress from self-interaction compared to the total contribution from each $k_y$ increases with increasing collisionality, while the second and third diagnostics reflecting that the combined effect of self-interaction from the multiple $k_y$s decreases with increasing collisionality. In section~\ref{SecCompdiffSIDiags}, a discussion is provided to understand this apparent contradiction.

\subsubsection{Normalised self-interaction contribution to Reynolds stress} \label{SecNormSIRS}
Reynolds stress, more exactly the off-diagonal component $\langle \tilde{V}_x\tilde{V}_\chi\rangle$ of the Reynolds stress tensor resulting from the combination of fluctuating $E\times B$ flow components in the radial and poloidal directions, can be considered as a proxy for the drive of zonal flows \cite{AjayCJ2020}. In particular, the radial conservation equation for the total gyrocenter charge density, in the electrostatic limit, invoking $m_e/m_i\ll 1$, and making use of the quasi-neutrality equation in the limit of long wavelength (correct to second order in $k_\perp\rho_i$), leads to an equation relating the shearing rate $\omega_{E\times B}$ associated to $E\times B$ zonal flows with the Reynolds stress RS:
	\begin{align}
	\frac{\partial }{\partial t} \omega_{E\times B}\sim \frac{\partial^2 }{\partial x^2} {\rm RS}.
	\label{EqwsimRS}
	\end{align}
Refer appendix B in Ref.~\onlinecite{AjayCJPhD} for the full equation and derivation.

  In GENE coordinates, this Reynolds stress can be defined as 
  \begin{equation}
    \label{EqRSdef}
    {\rm RS}(x,t)
    =
    \left\langle
    \frac{1}{B_0^2}\frac{\partial \Phi}{\partial y}
    \left(
    g^{xx}\frac{\partial \Phi}{\partial x}
    +
    g^{xy}\frac{\partial \Phi}{\partial y}
    \right)
    \right\rangle_{yz}.
  \end{equation}
Furthermore, one can write
  \begin{equation}
    \label{EqRS, sum}
    {\rm RS}(x, t) 
    =
    \sum_{k_y > 0} \hat{\rm RS}_{k_y}(x, t),
  \end{equation}
with the contribution from the $k_y$ Fourier mode given by
  \begin{equation}
    \label{RS_ky}
    \hat{\rm RS}_{k_y}(x, t)
    =
    2\,\text{Re}
    \left[\;
      \left\langle
      \frac{1}{B_0^2}\,k_y\hat\Phi_{k_y} 
      \left(
      g^{xx}{\rm i}\,\frac{\partial \hat\Phi_{k_y}^\star}{\partial x}
      +
      g^{xy} k_y \hat\Phi_{k_y}^\star
      \right)
      \right\rangle_z
    \right],
  \end{equation}
having invoked the reality condition $\hat\Phi_{-k_y}=\hat\Phi_{k_y}^\star$. Considering as well the $k_x$ Fourier mode decomposition of $\Phi$, each of these $k_y$ contributions can also be written as follows:
  \begin{align}
    \label{RS_ky in kx-repr}
    & \hat{\rm RS}_{k_y}(x, t)
    =
    \\
    &
    2\,\text{Re}
    \Bigg\{\;
    \sum_{k_x, k_x''}
    \left\langle
    \frac{1}{B_0^2}\,k_y
    \left(
    g^{xx} k_x''
    +
    g^{xy} k_y 
    \right)\hat\Phi_{k_x,k_y} \hat\Phi_{k_x'',k_y}^\star
    \right\rangle_z
    e^{{\rm i}(k_x-k_x'')x}
    \;\Bigg\},
  \end{align}
illustrating the drive of zonal modes $(k_x'=k_x-k_x'', 0)$ through non-linear interaction between Fourier modes $(k_x, k_y)$ and $(k_x'',k_y)$.

Given that the self-interaction drive of zonal flows results from the nonlinear interaction involving any two linearly coupled $k_x$-Fourier modes belonging to the same eigenmode [see Eq.~(\ref{linear eigenmode, Fourier rep.})], the  self-interaction contribution to Reynolds stress for a particular $k_y$ can be written as
  \begin{align}
    \label{RS_ky SI}
    & \hat{\rm RS}_{k_y}^{\rm si}(x, t)
    =
    2\,\text{Re}
    \Bigg\{\;
    \sum_{k_x} \hspace{0.2cm}
	\sum_{\substack{p\\ k_x''=k_x + p2\pi k_y\hat{s}}}
    \\
    & \left\langle
    \frac{1}{B_0^2}\,k_y
    \left(
    g^{xx} k_x''
    +
    g^{xy} k_y 
    \right)\hat\Phi_{k_x,k_y} \hat\Phi_{k_x'',k_y}^\star
    \right\rangle_z
    e^{{\rm i}(k_x-k_x'')x}
    \;\Bigg\},
  \end{align}
where $p$ is an integer index running from $-\infty$ to $\infty$ (or more practically, the maximum limits of the Fourier domain being considered in the simulation). Note that this definition of $\hat{\rm RS}_{k_y}^{\rm si}$ is equivalent to Eq.~(5.17) in Ref.~\onlinecite{AjayCJ2020}.

In the following analysis, one considers the double partial radial derivative of the total Reynolds stress, \emph{i.e.} $\partial^2\hat{\rm RS}_{k_y}/\partial x^2$, and the self-interaction contribution, \emph{i.e.} $\partial^2\hat{\rm RS}^{\rm si}_{k_y}/\partial x^2$, following their relation to zonal flow drive as given in Eq.~(\ref{EqwsimRS}). The normalised self-interaction contribution to Reynolds stress is now defined as the time average of the (the radial derivative of the) self interaction contribution to Reynolds stress normalised by the RMS over time of the (radial derivative of the) total Reynolds stress contribution, \emph{i.e.} $\mathcal{R} = \langle\partial^2\hat{\rm RS}^{\rm si}_{k_y}/\partial x^2\rangle_t/  {\rm RMS}(\partial^2\hat{\rm RS}_{k_y}/\partial x^2)$. This diagnostic simultaneously measures the relative importance of the self-interaction drive of zonal flows with respect to the total contribution to Reynolds stress drive from the considered $k_y$, as well as how good its sign is fixed at each radial position over time, which is a characteristic feature of the self-interaction mechanism. 

\begin{figure}
	\hspace{-0.8cm}
	\begin{minipage}{0.45\columnwidth}
	\includegraphics[width=1.5\textwidth] {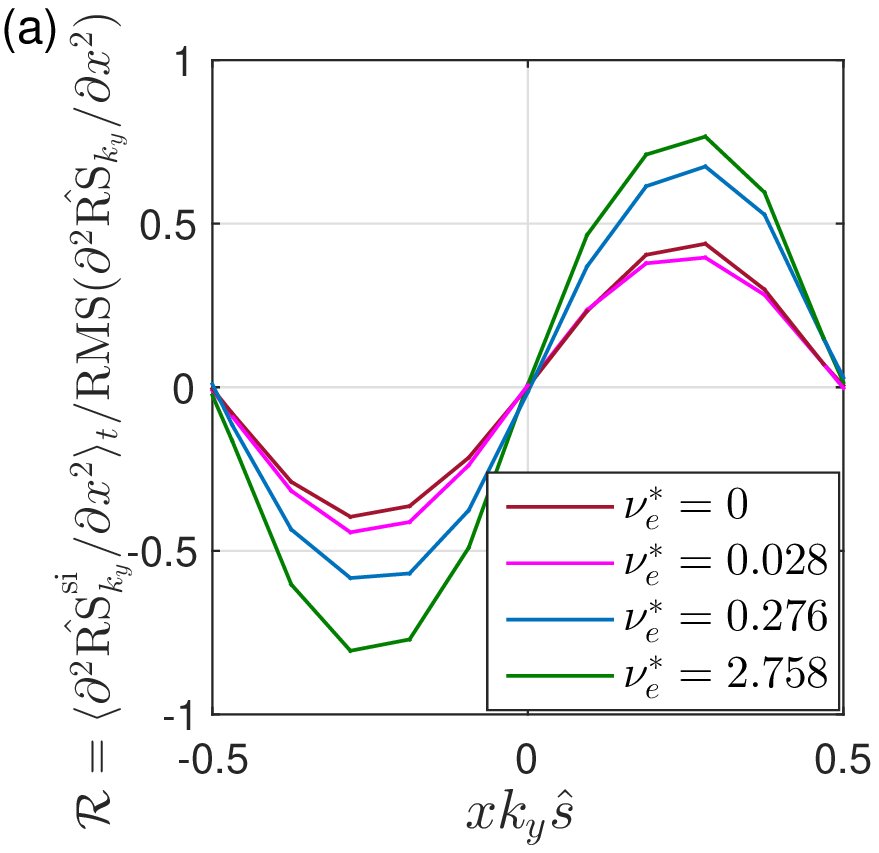}
	\end{minipage}
	\hspace{0.4cm}
	\begin{minipage}{0.45\columnwidth}
	\includegraphics[width=1.5\textwidth] {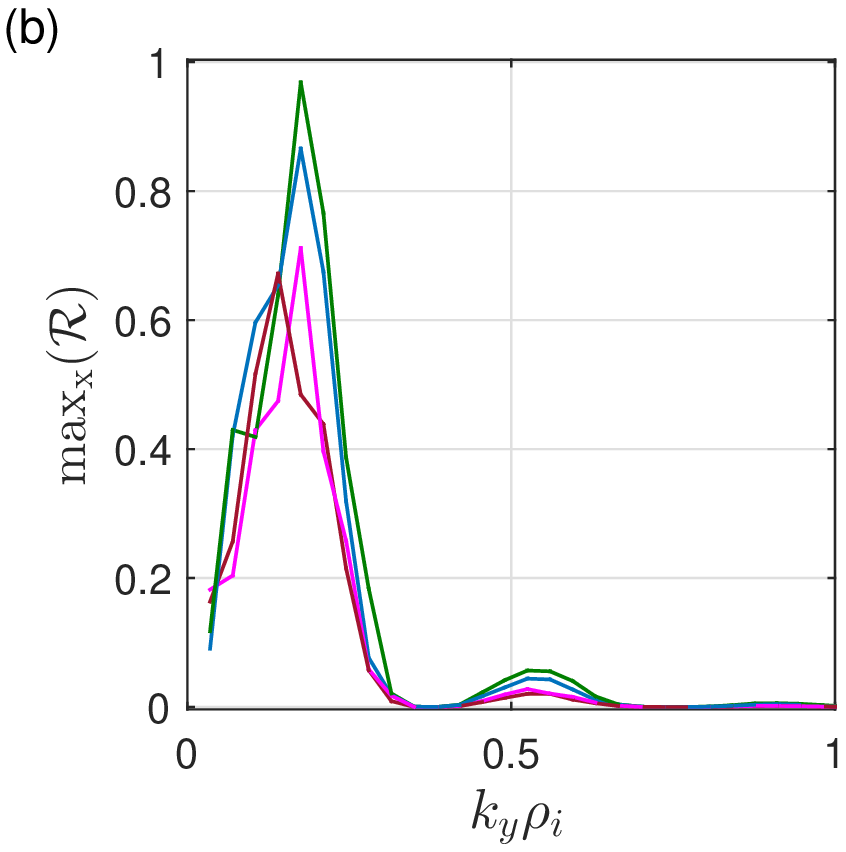}
	\end{minipage}
	\vspace{-1.0cm}
	\caption{(a) Time-average of the self-interacting contribution to Reynolds stress 
  normalised with respect to the RMS in time of the total contribution, \emph{i.e.} $\mathcal{R}=\langle\partial^2\hat{\rm RS}^{\rm si}_{k_y}/\partial x^2\rangle_t/  {\rm RMS}(\partial^2\hat{\rm RS}_{k_y}/\partial x^2)$, plotted as a function of the radial coordinate $x$ for $k_y\rho_i=0.21$. (b)  Maximum of $\mathcal{R}$ along the radial coordinate $x$, plotted as a function of $k_y\rho_i$. Results from turbulence simulations with collisionalities $\nu_e^*=0$ (brown), $0.028$ (magenta), $0.276$ (blue) and $2.758$ (green) are shown.}
	\label{d2RSsibyrms_collcomp}
	\end{figure}

In Fig.~\ref{d2RSsibyrms_collcomp}(a), $\mathcal{R}$ for $k_y\rho_i=0.21$ [contributing significantly to the $|\Phi|^2$ $k_y$-spectra, see Fig.~\ref{Phi2kyspectra_LTicomp}(a)] in turbulence simulations with the different considered collisionalities is plotted as a function of the radial coordinate $x$ over the interval $\Delta x_{\rm MRS}=1/k_y\hat{s}$ between MRSs, with $x=0$ the position of a corresponding MRS. As already discussed in section~\ref{SecCollNonlinFSWidth}, the nonlinear broadening mechanism radially widens these structures to essentially a sinusoid with a period $\Delta x_{\rm MRS}$. For the particular case of $k_y\rho_i=0.21$ considered in Fig.~\ref{d2RSsibyrms_collcomp}(a), it is found that the normalised measure $\mathcal{R}$ becomes more significant with increasing collisionality, with the exception of the collisionless case. To study the dependence of $\mathcal{R}$ on all $k_y$s, its maximum in $x$ is plotted as a function of $k_y$ in Fig.~\ref{d2RSsibyrms_collcomp}(b). The peak of these plots, measuring the maximum intensity of self-interaction as measured by the normalised self-interaction contribution to Reynolds stress $\mathcal{R}$, is found to increase with increasing collisionality. 

Note that, the more the linear characteristics of an eigenmode is retained in nonlinear simulations, greater will be the contribution to $\hat{\rm RS}^{\rm si}_{k_y}$ and the diagnostic $\mathcal{R}$. Hence the above result is consistent with the relative phase evolution analysis presented in Fig.~\ref{Phasediffvst_collcomp}, which shows an increased retention of the linear phase difference along the ballooning structure with increasing collisionality in nonlinear simulations.

\subsubsection{Bicoherence analysis}\label{SecBicoh}

Apart from the self-interaction mechanism, another, well-known zonal flow driving mechanism is that via the modulational instability mechanism~\cite{Hasegawa1978,Hasegawa1979,Chen2000}. As compared to the self-interaction mechanism, which leads to incoherent and random kicks to the zonal flow drive from each $k_y$ contribution~\cite{AjayCJ2020}, modulational instability is a coherent mechanism involving resonant 3-wave interactions, which in turn require frequency matching between the three participating Fourier modes. The strength of a particular resonant interaction between 3 Fourier modes, ${\bf k}=(k_x,k_y)$, the zonal mode ${\bf k'}=(k'_{x},0)$ and daughter mode ${\bf k''}={\bf k}-{\bf k'}=(k_x-k'_x,k_y)$, can be measured via a bicoherence-type analysis that is explained in detail below. Through this diagnostic, one aims to measure the relative significance of the self-interaction mechanism in driving zonal flows, over the modulational instability mechanism.

The bicoherence-type analysis essentially involves the time-average of the triplet product
    \begin{equation}
        T({\bf k}\ ;+{\bf k'})=\hat{\Phi}_{\bf k}(t)\hat{\Phi}^*_{\bf k'}(t)\hat{\Phi}^*_{\bf k''}(t)
    \end{equation}
where  $\hat{\Phi}_{\bf q} (t) \sim {\rm exp}[-i(\omega_{\bf q} t + \phi_{\bf q})]$ is the complex time dependent amplitude of the Fourier mode component ${\bf q}$ of the electrostatic field, having a frequency $\omega_{\bf q}$, phase shift $\phi_{\bf q}$, and evaluated at z=0. If the Fourier modes $[\mathbf k, \mathbf k', \mathbf k'' = \mathbf k-\mathbf k']$ are frequency matched, i.e. $\omega_{\bf k} = \omega_{\bf k'} + \omega_{\bf k''}$, then $\langle T({\bf k}\ ;+{\bf k'})\rangle_t\neq 0$, where $\langle .\rangle_t$ stands for the time-average over the simulation time. 

Now, a normalised measure of the strength of the resonant 3-wave interaction can be calculated by the following estimate: 
    \begin{equation}
        b_{N}({\bf k}\ ;+{\bf k'}) = \frac{|\langle T({\bf k}\ ;+{\bf k'})\rangle_t|}{\langle |T({\bf k}\ ;+{\bf k'})|\rangle_t},
    \end{equation}
defined as the bicoherence between the Fourier triplet $[\mathbf k, \mathbf k', \mathbf k'' = \mathbf k-\mathbf k']$. Note $0\leq b_{N}\leq 1$. Further, $b_{N}({\bf k}\ ;+{\bf k'})\simeq 1$ indicates a fully resonant 3-wave interaction between ${\bf k}$, ${\bf k''}$ and the zonal mode ${\bf k'}$, while $b_{N}\simeq 0$ indicates a non-resonant process. Since modulational instability is a simultaneous resonant interaction between both triplets $[\mathbf k, \mathbf k', \mathbf k'' = \mathbf k-\mathbf k']$ and $[\mathbf k, -\mathbf k', \mathbf k''' = \mathbf k+\mathbf k']$, we define total bicoherence
    \begin{equation}
        B_{N}({\bf k}\ ;{\bf k'}) = (b_N({\bf k}\ ;+{\bf k'})+b_N({\bf k}\ ;-{\bf k'}))/2,
        \label{Eq_bicoh}
    \end{equation} 
such that a value of $B_{N}\simeq 1$ indicates a fully resonant interaction characteristic of zonal flow drive dominated by the modulational instability mechanism. Values of $B_N$ closer to zero are however indicative of non-resonant interactions typical of zonal flow drive dominated by the self-interaction mechanism.

\begin{figure*}
	\centering
	\includegraphics[trim= 0cm 0cm 7cm 0cm, clip=true, width=1.8\columnwidth] {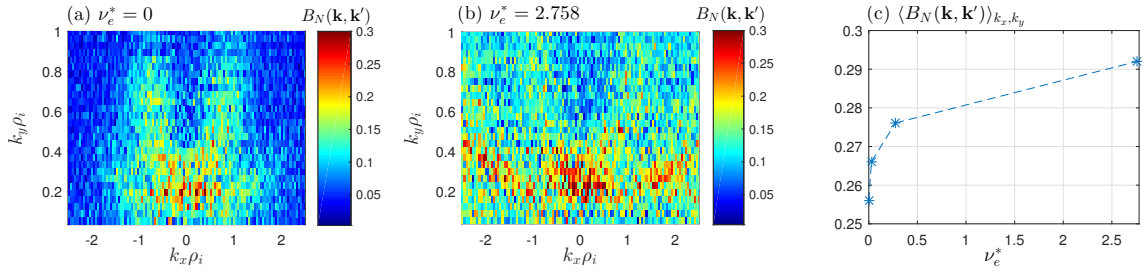}
	\vspace{-0.85cm}
	\caption{The bicoherence level $B_N({\bf k}=(k_x,k_y),{\bf k'})$ plotted as a function of $k_x$ and $k_y$, for the zonal mode ${\bf k'}=(k'_x,0)$ with $k'_x\rho_i=0.31$, in turbulence simulation with (a) no collisions and (b) collisionality $\nu_e^*=2.758$. (c) Average bicoherence level $\langle B_N\rangle_{k_x,k_y}$ plotted as a function of collisionality $\nu_e^*$.}
	\label{bicoh_collcomp}
	\end{figure*}

In Fig.~\ref{bicoh_collcomp}(a) and (b), $B_N({\bf k};{\bf k'})$ is plotted as a function of $k_x$ and $k_y$, for the zonal mode ${\bf k'}=(k'_x\rho_i=0.31,0)$ [which has a significant contribution to the $k_x$-spectra of effective zonal shearing rate $\omega_{\rm eff}$] in turbulence simulations with (a) no collisions and (b) collisionality $\nu_e^*=2.758$. Clearly, with collisions, the bicoherence levels are higher. To quantify the increase in the bicoherence levels with collisions, an average of $B_N({\bf k};{\bf k'})$ over $k_x$ and $k_y$, \emph{i.e.} $\langle B_N({\bf k};{\bf k'})\rangle_{k_x,k_y}$ as defined below in Eq.~(\ref{EqBicohAvg}), is plotted in Fig.~\ref{bicoh_collcomp}(c) as a function of $\nu_e^*$. 
\begin{align}
\label{EqBicohAvg}
\langle B_N({\bf k};{\bf k'})\rangle_{k_x,k_y} 
= \frac{k_{x,{\rm min}} k_{y,{\rm min}}}{\Delta k^2} \sum\limits_{k_x,k_y} B_N({\bf k}=(k_x,k_y);{\bf k'}),
\end{align}
where
\begin{align}
\nonumber
\Delta k^2
&= \frac{\sum\limits_{k_x,k_y} [(k_x-k_{x,{\rm av}})^2 + (k_y-k_{y,{\rm av}})^2]\  |B_N({\bf k}=(k_x,k_y);{\bf k'})|}{\sum\limits_{k_x,k_y} |B_N({\bf k}=(k_x,k_y);{\bf k'})| }  
\\
\nonumber
{\rm and}
\\
\nonumber
k_{\alpha,{\rm av}}
&= \frac{\sum\limits_{k_x,k_y} k_{\alpha} |B_N({\bf k}=(k_x,k_y);{\bf k'})| }{\sum\limits_{k_x,k_y} |B_N({\bf k}=(k_x,k_y);{\bf k'})| }
\end{align}
with $k_{\alpha,{\rm av}}=k_{x,{\rm av}},k_{y,{\rm av}}$. Note that this average of $B_N({\bf k}=(k_x,k_y);{\bf k'})$
 as defined above is independent of $N_{k_x}$ (\emph{i.e.} $k_{x,{\rm max}}$) and $N_{k_y}$ (\emph{i.e.} $k_{y,{\rm max}}$), which would not be the case if one were to consider a simple average of the form $\sum_{k_x,k_y}B_N({\bf k}=(k_x,k_y);{\bf k'})/N_{k_x}N_{k_y}$. As already mentioned, higher bicoherence levels are characteristic of increased zonal flow drive from the modulational instability mechanism, whereas the self-interaction contribution to Reynolds stress from the different $k_y$s, being uncorrelated with each other and random in time, lead to lower levels of bicoherence. Hence, the increase in $\langle B_N({\bf k};{\bf k'})\rangle_{k_x,k_y}$ with collisionality suggests that collisions weaken the self-interaction mechanism. 

In the following, the analysis based on correlation between the various $k_y$ contributions to Reynolds stress is discussed. 

\subsubsection{Correlation between the various $k_y$ contributions to Reynolds stress}\label{SecCorrRS}

As already mentioned before, the self-interaction mechanism leads to incoherent and random kicks to the zonal flow drive from each $k_y$ contribution, whereas modulational instability leads to more correlated drive. To further quantify this, one defines an effective correlation function ${\rm C_{RS}}$ measuring the average correlation between all pairs of [$\partial^2\hat{\rm RS}_{k_{y,i}}/\partial x^2$, $\partial^2\hat{\rm RS}_{k_{y,j}}/\partial x^2$] for $k_{y,i}\neq k_{y,j}$:
    \begin{equation}    
    {\rm C_{RS}}[f]= \sum_{\substack{k_{y,i},\ k_{y,j} \\ k_{y,j}>k_{y,i}}
    }\frac{{\rm Cov}[\hat{f}_{k_{y,i}},\hat{f}_{k_{y,j}}]}{\sigma[\hat{f}_{k_{y,i}}]\sigma[\hat{f}_{k_{y,j}}]} \ \Big/
    \sum_{\substack{k_{y,i},\ k_{y,j} \\ k_{y,j}>k_{y,i}}} 1\ .
    \label{Eq_normcorr}
    \end{equation}
$f=\partial^2{\rm RS}/\partial x^2$, $\hat{f}_{k_{y}}$= $\partial^2\hat{\rm RS}_{k_y}(x)/\partial x^2$ as defined in Eq.~(\ref{RS_ky}), covariance Cov$[a,b]=(\sigma^2[a+b]-\sigma^2[a]-\sigma^2[b])/2$, and variance $\sigma^2[a]=\langle|a-\langle a\rangle_t|^2\rangle_t$, with $\langle .\rangle_t$ representing average over simulation time.  Note that ${\rm C_{RS}}\in [0,1]$, with 1 corresponding to perfect correlation between Reynolds stress drive from all $k_y$'s and 0 corresponding to total decorrelation between them.

\begin{figure}
	\centering
	\includegraphics[width=1.25\columnwidth] {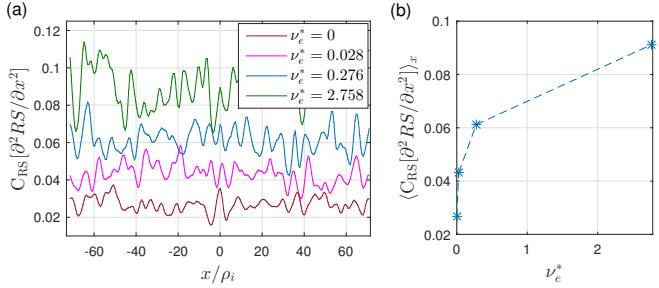}
	\vspace{-1.3cm}
	\caption{(a) Correlation ${\rm C_{RS}}$ between the $k_y$ modes of Reynolds stress contributions $\partial^2{\rm RS}/\partial x^2$ as a function of $x$ in turbulence simulations with collisionalities $\nu_e^*=0$ (brown), $0.028$ (magenta), $0.276$ (blue) and $2.758$ (green). (b) Radial average of the correlation $\langle{\rm C_{RS}}[\partial^2{\rm RS}/\partial x^2]\rangle_x$ plotted as a function of collisionality $\nu_e^*$.}
	\label{SumNormCorr_collcomp}
	\end{figure}

In Fig.~\ref{SumNormCorr_collcomp}(a), the normalised correlation ${\rm C_{RS}}[\partial^2{\rm RS}/\partial x^2]$ between the $k_y$ modes of Reynolds stress contributions is plotted as a function of $x$ for turbulence simulations with different collisionalities. And in Fig.~\ref{SumNormCorr_collcomp}(b), the corresponding radial average $\langle{\rm C_{RS}}[\partial^2{\rm RS}/\partial x^2]\rangle_x$ is plotted as a function of $\nu_e^*$. It is found that the normalised correlation between the $k_y$ modes of Reynolds stress contributions, as measured by ${\rm C_{RS}}[\partial^2{\rm RS}/\partial x^2]$, increases with increasing collisionality. This is consistent with the conclusion based on the bicoherence analysis, that collisions weaken the (incoherent) self-interaction drive mechanism and lead to a relative dominance of the (coherent) modulational instability mechanism.

However these results are seemingly in contradiction to the result from the first diagnostic that measures the effect of self-interaction independently for each $k_y$. In the following, an attempt is made to resolve this apparent contradiction.

\subsubsection{Comparing diagnostics measuring the significance of self-interaction from each $k_y$ separately and from all $k_y$s simultaneously}\label{SecCompdiffSIDiags}

The first diagnostic, measuring the normalised self-interaction contribution to Reynolds stress (in Fig.~\ref{d2RSsibyrms_collcomp}), estimates the importance of the self-interaction contribution independently for each $k_y$. The result from this diagnostic shows that with increasing collisionality, the eigenmodes retain more of their linear characteristics in turbulence simulations, consistent with the linear phase difference evolution analysis in Fig.~\ref{Phasediffvst_collcomp}. Given that higher RMS amplitudes of physical quantities in turbulence simulations is indicative of a system being more nonlinear, it is consistent to expect simulations with higher collisionalities, which consequently have less unstable linear eigenmodes, and therefore reduced RMS amplitudes of physical quantities, to be less nonlinear. In the $k_y$ spectra of $|\Phi|^2$ in Fig.~\ref{Phi2kyspectra_LTicomp}(a) for each of the four turbulence simulations considered in this paper, one can indeed see that the RMS amplitudes and therefore the overall level of nonlinearity is lower in simulations with higher collisionalities.

On the other hand, the the bicoherence analysis (in Fig.~\ref{bicoh_collcomp}) and the analysis based on the correlation between the different $k_y$ contributions to Reynolds stress (in Fig.~\ref{SumNormCorr_collcomp}), measures the collective effect of self-interaction from multiple $k_y$s, \emph{i.e.}, these diagnostics also account for how the different $k_y$s compete with each other to drive zonal flows via self-interaction.

While the first diagnostic indicates that the relative self-interaction contribution from each $k_y$ increases with increasing collisionality, the second and third diagnostics reflect that the combined effect of self-interaction from the multiple $k_y$s decreases with increasing collisionality. These different results may \emph{a priori} appear contradictory. However, it should be noted that the number of significant $k_y$ modes participating in turbulence and how nonlinear the system is plays a significant role in determining the total effect of self-interaction. This can be explained using the following thought experiment: Consider a case where the linear eigenmode continues to become less unstable as one increases collisionality (Note that in reality, this is true only until electron response becomes fully adiabatic, beyond which the growth rate plateaus as shown in Fig.~\ref{gammaandomegavscoll}). Now, consider a nonlinear system with high enough collisionality such that effectively the microturbulence is driven by a single unstable eigenmode having a particular $k_y$. In this limit of just one mode contributing to Reynolds stress via self-interaction, corresponding kicks always drive the zonal flows in the same direction at a given radial position. Whereas with multiple $k_y$s contributing, as would in a system with low collisionality, the Reynolds stress kicks to zonal flows from different $k_y$s have different signs at any particular radial position (since the MRSs of each $k_y$ are mis-aligned in the radial coordinate), and given that they are furthermore uncorrelated in time, act as random kicks that more effectively disrupt the coherent drive from modulational instability. 

One thus concludes that in case with high collisionality, where the system is less nonlinear, the bicoherence analysis and the correlation between the various $k_y$ contributions to Reynolds stress show a decrease in the total effect of self-interaction.

\subsubsection*{Parallel between increasing collisionality and decreasing $R/L_{T,i}$}
\addcontentsline{toc}{subsubsection}{\small - Parallel between increasing collisionality and decreasing $R/L_{T,i}$}

\begin{figure}
\centering
  \includegraphics[width=1.02\columnwidth] {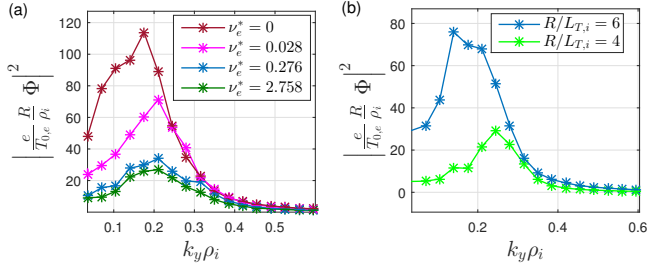}
  \caption{(a) $k_y$ spectra of $|\Phi|^2$ in turbulence simulations whose parameters are given in table~\ref{CollParameterSet}, with collisionalities $\nu_e^*=0$ (brown), $0.028$ (magenta), $0.276$ (blue) and $2.758$ (green). (b) $k_y$ spectra of $|\Phi|^2$ for the collisionless kinetic electron turbulence simulations given in Ref.~\onlinecite{AjayCJ2020}, with $R/L_{T,i}=6$ (blue) and 4 (green).}
  \label{Phi2kyspectra_LTicomp}
\end{figure}

A clear illustration of a reduced number of $k_y$ modes contributing to turbulence and self-interaction is observed when the background ion temperature gradient is decreased. In Fig.~\ref{Phi2kyspectra_LTicomp}(b), the $k_y$ spectra of $|\Phi|^2$ of two such simulations are shown, with $R/L_{T,i}=6$ and $4$, with the latter being close to marginal stability. These are the same set of kinetic electron simulations presented in Ref.~\onlinecite{AjayCJ2020}, having parameters similar to that in table~\ref{CollParameterSet} but with different background gradients, a mass ratio of $m_i/m_e=400$ and slightly different numerical resolutions.

\begin{figure}
\centering
  \includegraphics[width=0.75\columnwidth] {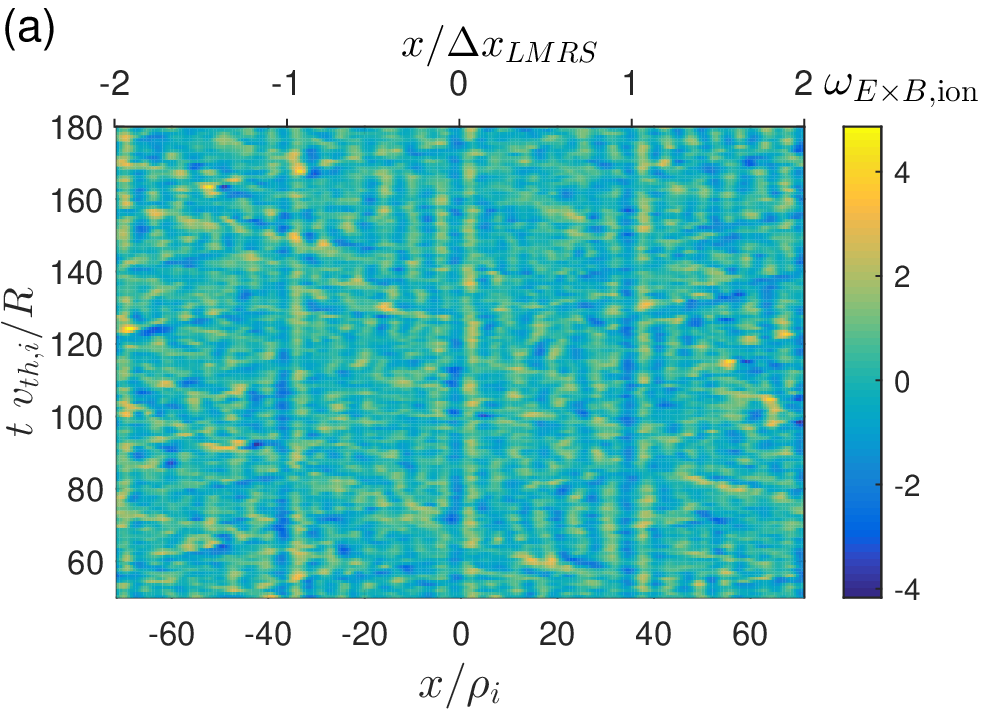}
  \includegraphics[width=0.75\columnwidth] {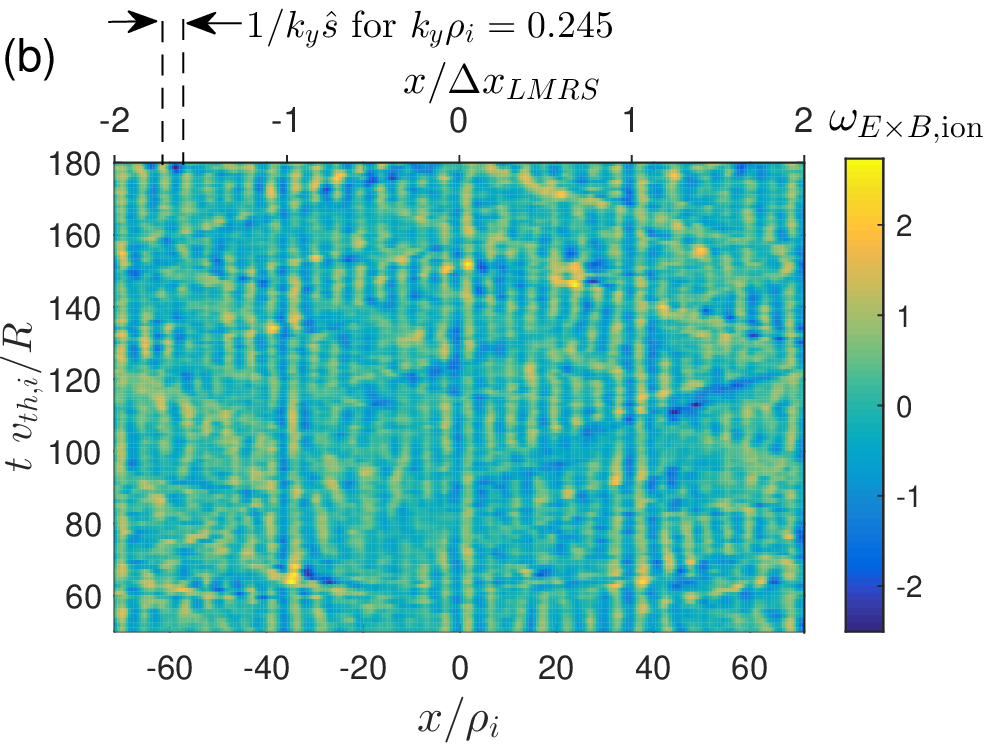}
  \caption{$\omega_{E\times B,ion}$ as a function of $x$ and $t$ in simulations with (a) $R/L_{T,i}=6$  and (b) 4. Other parameters are the same as in table~\ref{CollParameterSet}.}
  \label{wonxandt_LTicomp}
\end{figure}

For the case with $R/L_{T,i}=4$, one can clearly observe that $k_y\rho_i=0.245$ contributes a large fraction of the total fluctuation energy. Therefore, in the corresponding plot of zonal flow shearing rate $\omega_{E\times B,{\rm ion}}$ shown in Fig.~\ref{wonxandt_LTicomp}(b), one can see significant stationary structures driven by self-interaction at the corresponding MRSs separated by a distance $\Delta x_{\rm MRS}=1/\hat{s}k_y=5.10\rho_i$. Whereas in the case far from marginal stability, \emph{i.e.} for $R/L_{T,i}=6$ in Fig.~\ref{wonxandt_LTicomp}(a), the stationary self-interaction contributions from the larger number of $k_y$s, being radially mis-aligned, tend to cancel each other out on average between lowest order MRSs. 



In the collisionless simulations with $R/L_{T,i}=6$ and $4$, the average bicoherence level $\langle B_N({\bf k};{\bf k'})\rangle_{k_x,k_y}$ are 0.113 and 0.160 respectively. The corresponding correlation levels $\langle{\rm C_{RS}}[\partial^2{\rm RS}/\partial x^2]\rangle_x$ are 0.006 and 0.011 respectively. Both these diagnostics indicate that the total effect of self-interaction decreases as one moves closer to marginal stability. This provides further validation to the hypothesis that self-interaction is less disruptive to modulational instability in cases nearer to marginal stability, for which a reduced number of $k_y$ modes contribute significantly to the turbulence drive.

\section{Conclusions}\label{SecCollConclusions}
The effect of collisions on the non-adiabatic passing electron dynamics has been studied in this work using both linear and nonlinear gyrokinetic simulations. In linear simulations, the weakening of the non-adiabatic electron drive of ITG microinstability with collisions, which has already been reported in Ref.~\onlinecite{Mikkelsen2008}, has been shown to be a consequence of the increased adiabatic like response of electrons away from MRSs. In addition, it is found that the characteristic parallel length scale associated to the ballooning envelope tail of the eigenmodes is set primarily by the electron-ion mean free path. This in turn leads to an increase in the radial width of the fine-structures with increasing collisionality. 

The decrease in the linear drive of the microinstability with increasing collisions leads to a corresponding decrease in the heat and particle flux levels in nonlinear simulations. The radial width of fine-structures in nonlinear simulations is found to be set predominantly by the nonlinear broadening mechanism. As a result, a slight decrease in the radial width with increasing collisionality is observed in turbulence simulations. 

Finally, the effect of collisions on the self-interaction mechanism is studied using three diagnostics, the first measuring the effect of self-interaction independently for each $k_y$ and the second and third measuring the total effect of self-interaction simultaneously for multiple $k_y$s. While the first diagnostic indicates that  the self-interaction for each $k_y$ increases with increasing collisionality, the other two show that the total effect of self-interaction in a nonlinear simulation decreases with increasing collisionality. This is explained by the decreased nonlinearity and the associated lower number of significant $k_y$s contributing towards self-interaction in simulations with higher collisionality where the flux and amplitude levels of physical quantities are also lower.

The final take-away from this paper is that for physically relevant values of collisionality in the core, the effect of non-adiabatic passing electrons, in particular the self-interaction mechanism, remains significant.

\begin{acknowledgments}
This work has been carried out within the framework of the EUROfusion Consortium and has received funding from the Euratom research and training programme 2014 - 2018 and 2019 - 2020 under grant agreement No 633053. The views and opinions expressed herein do not necessarily reflect those of the European Commission. We acknowledge the CINECA award under the ISCRA initiative, for the availability of high performance computing resources and support. Lastly, this work was supported by a grant from the Swiss National Supercomputing Centre (CSCS) under project ID s956 and s1050.
\end{acknowledgments}

\section*{Data Availability Statement}
The data that support the findings of this study are available from the corresponding author upon reasonable request.

\section*{References}
\vspace{-0.3cm}
\bibliography{ch_Collisions}

\end{document}